\documentclass[]{pasj02}
\Received{}%{yyyy/mm/dd}
\Accepted{}%{yyyy/mm/dd}
\usepackage{booktabs}
\usepackage[switch,mathlines]{lineno}
\begin{document} 

\title{
XRISM Resolve Spectroscopy of GX~5--1: Constraints on Iron Spectral Features in a Luminous Neutron-Star Binary
}
%%% begin:list of authors
% Do NOT capitalize all letters in "textsc".
\author{Tasuku Hayashi\altaffilmark{1,2}\thefootnote{*}\orcid{0000-0002-6587-9314}}
\email{tasuku.hayashi@riken.jp}
\author{Shinya Yamada\altaffilmark{2}\orcid{0000-0003-4808-893X}} 
\author{Shun Inoue\altaffilmark{3}}
\author{Teruaki Enoto\altaffilmark{3}}
\author{Noriko Yamasaki\altaffilmark{4}}
\author{Shunji Kitamoto\altaffilmark{2}\orcid{0000-0001-8948-7983}}
\author{Shogo Kobayashi\altaffilmark{2}\orcid{0000-0001-7773-9266}}
\author{Yusuke Sakai\altaffilmark{2}\orcid{0000-0002-5809-3516}}
\author{Shintaro Kaneko\altaffilmark{2}\orcid{0009-0009-0927-0772}}
\author{Rin Ebisawa\altaffilmark{2}}
\author{Haruto Aoki\altaffilmark{2}}
\author{Kota Miyawaki\altaffilmark{2}}
\author{Misaki Mizumoto\altaffilmark{5}\orcid{0000-0003-2161-0361}}
\author{Hiromitsu Takahashi\altaffilmark{6,7}\orcid{0000-0001-6314-5897}}
\altaffiltext{1}{RIKEN Nishina Center for Accelerator-Based Science (RNC), 2-1 Hirosawa, Wako, Saitama 351-0198, Japan}
\altaffiltext{2}{Department of Physics, Rikkyo University, 3-34-1 Nishi Ikebukuro, Toshima-ku, Tokyo 171-8501, Japan}
\altaffiltext{3}{Department of Physics, Kyoto University, Kitashirakawa Oiwake, Sakyo, Kyoto 606-8502, Japan}
\altaffiltext{4}{Institute of Space and Astronautical Science (ISAS), Japan Aerospace Exploration Agency (JAXA), Kanagawa 252-5210, Japan} % ISAS/JAXA
\altaffiltext{5}{Science Research Education Unit, University of Teacher Education Fukuoka, Fukuoka 811-4192, Japan}
\altaffiltext{6}{Hiroshima Astrophysical Science Center, Hiroshima University, Hiroshima 739-8526, Japan} % Hiroshima ASC
\altaffiltext{7}{Department of Physics, Hiroshima University, Hiroshima 739-8526, Japan} % Hiroshima U
%% `\KeyWords{}' always has to be placed before ``\maketitle'' 
%%  List of Key Words:  https://academic.oup.com/pasj/pages/Pasj_Keywords 

% \noindent IMPORTANT NOTICE\\
% 1. ``\verb|\draft|'' creates single column and double spaces format.\\
% 2. If you comment out ``\verb|\draft|'', the output will be double column
%    and single space.\\
% 3. For cross-references, the use of ``\verb|\label|, \verb|\ref|, \verb|\cite|'' 
%    and the thebibliography environment is strongly recommended. \\
% 4. Do NOT use ``\verb|\def|, \verb|\renewcommand|''.\\
% 5. Do NOT redefine commands provided by PASJ01.cls.\\

% \newpage

\maketitle

 %%%
 \begin{abstract}

GX~5--1 is one of the brightest persistent neutron-star low-mass X-ray binaries in the Galaxy and one of the classical Z sources accreting at luminosities close to or above the Eddington limit.
For several decades, GX~5--1 has been characterized primarily by strong branch-dependent continuum and timing variability, while discrete Fe~K spectral signatures have remained weak or elusive compared with those in many other luminous neutron-star systems.
We present the first high-resolution X-ray spectroscopic study of GX~5--1 with the Resolve micro-calorimeter onboard the X-Ray Imaging and Spectroscopy Mission (XRISM), based on a 50~ks AO1 observation.
With an energy resolution of $\sim$5~eV at 6~keV, Resolve enables a sensitive non-dispersive search for weak and narrow spectral features in this luminous source.
During the observation, GX~5--1 was located on the horizontal branch of the Z track.
Although rapid spectral variability is detected on timescales of tens of ks, the variability is dominated by changes in the continuum spectral shape.
No statistically significant emission or absorption features are detected in the time-averaged spectrum.
Using matched-filter searches and time-resolved spectroscopy, we place stringent upper limits on the equivalent widths of low-ionization Fe~K$\alpha$ emission and highly ionized Fe~\textsc{xxv} and Fe~\textsc{xxvi} K$\alpha$ transitions of $EW \lesssim$ a few ~eV.
These results reinforce the picture that, in GX~5--1, classical Fe~K diagnostics such as reflection- or wind-related features are either intrinsically weak, strongly ionized, geometrically diluted, or suppressed by rapid continuum variability.
Placed in the context of the long observational history of GX~5--1 as a continuum-dominated Z source, the present XRISM result shows that this picture continues to hold even at calorimeter resolution.
GX~5--1 therefore provides an important benchmark for high-resolution studies of luminous neutron-star accretion flows and demonstrates the capability of XRISM to place meaningful constraints on line formation even when no significant features are detected.

\end{abstract}

%============================================================
\section{Introduction}
\label{sec:intro}

GX~5--1 is one of the brightest persistent neutron-star low-mass X-ray binaries (NS--LMXBs) in the Milky Way and is widely recognized as one of the classical Z sources accreting at luminosities close to, and possibly exceeding, the Eddington limit \citep{Hasinger1989-ab}.
Z sources trace characteristic Z-shaped tracks in color--color and hardness--intensity diagrams, reflecting correlated spectral and timing evolution under near-Eddington conditions.
Together with GX~340$+$0 and Cyg~X--2, GX~5--1 occupies the high-luminosity end of the NS--LMXB population and has long served as a benchmark system for studies of radiation-pressure-dominated accretion flows.

The spectral complexity of luminous NS--LMXBs was already recognized in the early era of X-ray spectroscopy.
Using \textit{Tenma}, \citet{Mitsuda1984-ox} showed that the spectra of bright NS--LMXBs can be described by a combination of multi-temperature disk emission and blackbody-like radiation, naturally interpreted as emission from an optically thick accretion disk and a neutron-star boundary layer.
This framework provided the foundation for subsequent interpretations of spectral evolution along the Z track.
In parallel, early high-time-resolution observations with the X-ray astronomy satellite \textit{GINGA} revealed quasi-periodic oscillations (QPOs) in GX~5--1, including horizontal-branch oscillations (HBOs), and established the intimate coupling between timing and spectral states \citep{Mitsuda1989-tx,Mitsuda1991-hl}.
Detailed cross-spectral and time-lag analyses further quantified the energy-dependent delays associated with HBOs and low-frequency noise, implying an important role for Comptonization and dynamical coronal processes \citep{Vaughan1994-sk}.

Comprehensive timing and spectral studies during the 1990s solidified GX~5--1 as a prototypical Z source.
Correlated spectral and timing behavior along the Z track was examined in detail by \citet{Kuulkers1994-zm} and \citet{Asai1994-zm}, while longer-term variability and secular changes were investigated by \citet{Kamado1997-zm}.
These studies showed that GX~5--1 is a source whose branch-dependent behavior is governed primarily by continuum and timing changes, a picture that has remained central to its interpretation.
Subsequent theoretical and phenomenological work connected the Z-track behavior to changes in mass accretion rate, inner disk structure, and the response of the boundary-layer region \citep{Jackson-N-K2009-jb,Sriram2011-mg}.

More recent observations with \textit{AstroSat} have revisited GX~5--1 with broad-band spectral and timing coverage across the Z track.
Anticorrelated soft--hard lags on timescales of tens to hundreds of seconds were reported by \citet{P2022-hi}, who constrained the size of the Comptonizing region to scales of tens of kilometers.
Comprehensive spectral--temporal analyses \citep{Thomas2024-zy,Shyam-Prakash2024-rg} have reinforced the view that much of the spectral evolution is driven by changes in the Comptonizing plasma and boundary-layer properties, while the inner disk temperature remains comparatively stable.
In addition, episodes of extended flaring have revealed radiative recombination continua (RRC) near 8--9~keV, interpreted as signatures of rapid expansion and cooling of a hot, thermalized boundary layer \citep{Dutta2024-da}.
Taken together, these results emphasize the strongly time-dependent and multi-component nature of the accretion environment in GX~5--1.

Despite its extreme luminosity, however, GX~5--1 has proven remarkably deficient in discrete Fe~K reflection signatures.
Broad-band \textit{NuSTAR} spectroscopy found no compelling evidence for relativistically broadened Fe~K$\alpha$ emission or a prominent Compton reflection hump \citep{Homan2018-ym}.
That result was especially noteworthy because many luminous NS--LMXBs do show such features.
The absence of reflection signatures in GX~5--1 was interpreted as evidence for a highly ionized inner disk, or alternatively for a geometry in which reflection is diluted or suppressed by the structure of the innermost accretion flow.
This raised an important question: whether GX~5--1 is intrinsically poor in observable Fe~K diagnostics, or whether weak narrow features have simply remained below the sensitivity of moderate-resolution instruments.
Previous high-resolution observations of neutron-star low-mass X-ray binaries have typically detected Fe~K emission or absorption features with equivalent widths of order 10--50~eV, placing practical sensitivity limits at the level of several tens of eV (e.g., \citep{Ueda2004-yg}; \citep{Miller2006-jm}; \citep{Ponti2012-gp}).

GX~5--1 has also played an important role in studies of the interstellar medium.
Energy-resolved X-ray halo measurements \citep{Smith2006-wn,Clark2018-by} and detailed modeling of the silicon K-edge fine structure \citep{Zeegers-S-T2017-ir}
have provided constraints on dust composition, scattering, and absorption along the line of sight.
More recently, variable energy-dependent X-ray polarization has been reported from GX~5--1 \citep{Fabiani2024-jc}, opening a new diagnostic window on the geometry and radiative processes of the inner accretion flow.
In many of these studies, GX~5--1 has effectively served as a bright and comparatively smooth continuum source, again suggesting that intrinsic narrow spectral features are weak or absent.

XRISM \citep{Tashiro2020-iz,Tashiro2025-ja} carries two co-aligned instruments, Resolve \citep{Ishisaki2022-ci,Sato2023-gr,Ishisaki2025-aw,Kelley2025-pz} and Xtend \citep{Mori2022-wo,Noda2025-aq,Uchida2025-ld}, at the focal planes of two X-ray Mirror Assemblies \citep{Tamura2024-wb}.
Resolve enables non-dispersive high-resolution spectroscopy of bright sources, with an energy resolution of $\sim$5~eV (FWHM) at 6~keV.
In the Fe~K band, this capability allows sensitive searches for weak and narrow emission or absorption features, including highly ionized iron transitions, narrow reflection components, disk-wind signatures, and fine structure around absorption edges.
Even in the absence of strong detections, stringent upper limits can place meaningful constraints on the ionization state, geometry, and covering fraction of the innermost accretion flow.

In this paper, we present the first XRISM/Resolve spectroscopic study of GX~5--1.
Rather than treating XRISM as a completely new starting point, we place the present observation in the context of the extensive observational legacy built by earlier missions.
Our goals are to use high-resolution spectroscopy in the Fe~K band (1) to test whether narrow Fe~K features are present at a level beyond the reach of previous instruments, (2) to quantify upper limits in the context of the reflection-deficient picture suggested by \cite{Homan2018-ym}, and (3) to relate the resulting constraints to the broader phenomenology of GX~5--1 as a classical Z source with a complex, variable boundary-layer and Comptonizing environment \citep{Dutta2024-da, Thomas2024-zy, Shyam-Prakash2024-rg}.
Through this approach, we aim to clarify whether GX~5--1 represents a genuinely Fe~K-suppressed accretion regime or a case in which subtle spectral diagnostics have remained hidden in previous observations.

%============================================================
\section{Observations and Data Reduction}
\label{sec:obs}

GX~5--1 was observed with XRISM during AO1 (ObsID~201060010) from 2025 April 15 13:47:04 to April 16 15:55:04 (UT), resulting in a total net exposure of $\sim50$~ks.
Data reduction was performed using the latest available calibration at the time of analysis (pre-pipeline 005\_003.20Jun2024\_Build8.014, pipeline 03.00.013.010, CALDB gen20241115\_xtd20241115\_rsl20241115).
Standard screening criteria were applied, excluding intervals affected by enhanced particle background, South Atlantic Anomaly (SAA) passages, telemetry saturation, Earth occultation, and bright limb contamination.

%------------------------------------------------------------
\subsection{Resolve}
\label{subsec:resolve}

Given the extreme source brightness, Resolve was operated with the neutral density (ND) filter to prevent pulse-shape distortion and event grading systematics at high count rates.
The ND filter consists of a 0.25~mm-thick molybdenum plate with 1.1~mm-diameter perforations \cite{de-Vries2017-oo, Shipman2024-tr}.
Although this configuration reduces the effective area, it preserves spectral fidelity in the Fe~K band, which is the primary focus of this study.
The gate valve remained in the closed configuration throughout the observation.
No adiabatic demagnetization refrigerator (ADR) recharging cycle \citep{Shirron2018-wf} occurred during the observation, resulting in uninterrupted exposure.

\begin{figure*}[h]
  \begin{center}
    \includegraphics[width=0.9\linewidth]{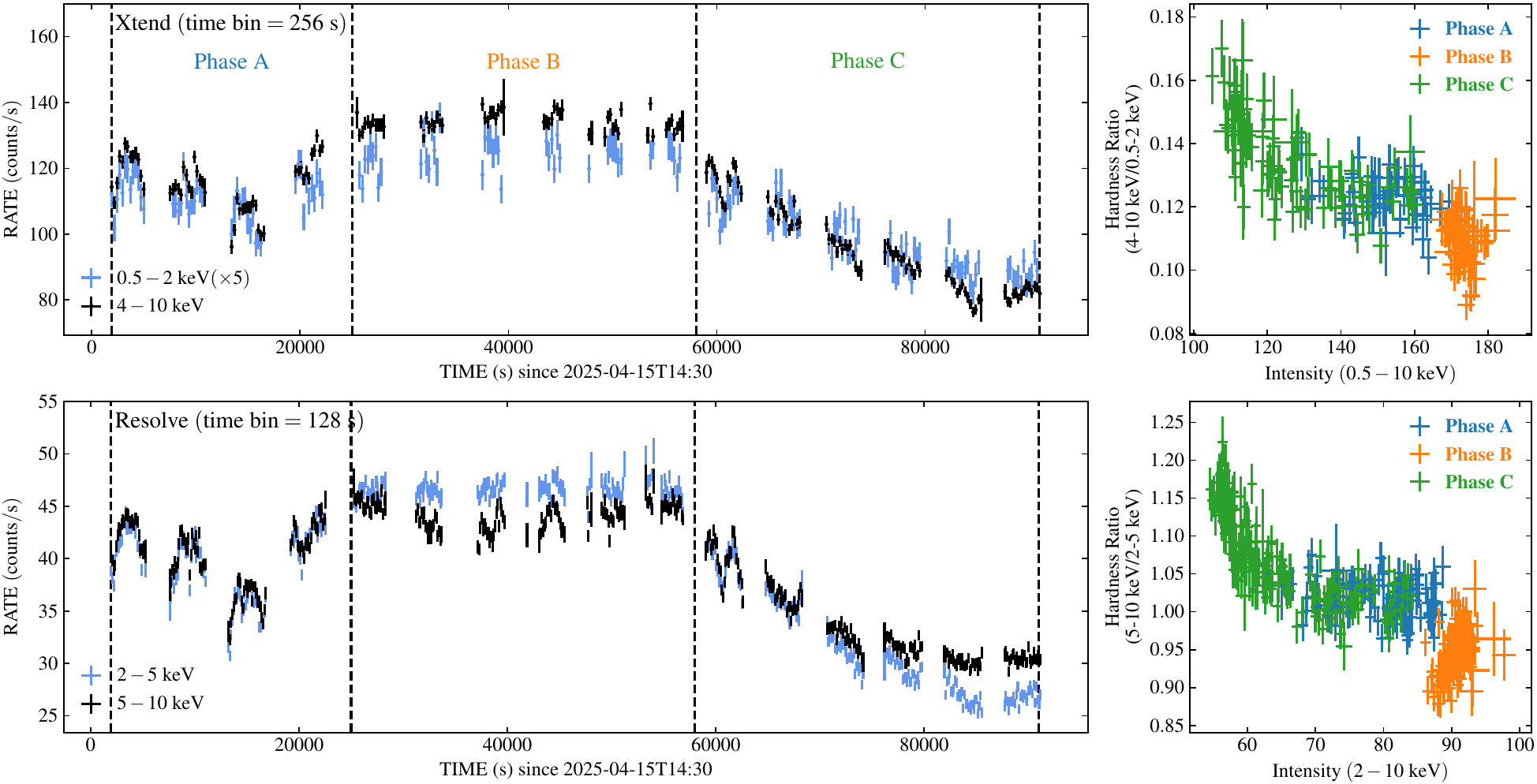}
  \end{center}
\caption{
(Left) Light curves of GX~5--1 obtained with XRISM.
The top-left panel shows the Xtend count rates in the 0.5--2~keV (blue; scaled by a factor of five for clarity) and 4--10~keV (black) bands with 256~s time bins.
The bottom-left panel presents the Resolve count rates in the 2--5~keV (blue) and 5--10~keV (black) bands with 128~s time bins.
The time axis is given in seconds since 2025 April 15 14:30 (UTC).
Based on the evolution of intensity and spectral hardness, the observation is divided into three intervals: Phase~A ($\sim$13.8~ks), Phase~B ($\sim$15.9~ks), and Phase~C ($\sim$21.4~ks), indicated by vertical dashed lines.
(Right) Hardness--intensity diagrams derived from the Xtend (top) and Resolve (bottom) data.
The hardness ratios are defined as (4--10~keV)/(0.5--2~keV) for Xtend and (5--10~keV)/(2--5~keV) for Resolve.
The hardness ratios are plotted against the total count rates in 0.5--10~keV (Xtend) and 2--18~keV (Resolve), respectively.
Data points are color-coded by phase (Phase~A: blue; Phase~B: orange; Phase~C: green).
The source exhibits pronounced intensity variations accompanied by systematic spectral evolution across the three phases.
{Alt text: XRISM observations of GX 5-1 showing light curves (Xtend and Resolve) divided into three phases based on intensity and spectral hardness evolution, and corresponding hardness-intensity diagrams color-coded by phase. The source exhibits large intensity variations accompanied by systematic spectral changes across the three phases.}
}
  \label{fig:lc}
\end{figure*}

The Resolve signal-processing chain, inherited from ASTRO-H SXS \citep{Kelley2016-xk}, includes the detector array, Xbox electronics (\cite{Porter2010-ae}; \cite{Porter2018-ip}), and the pulse shape processor (PSP; \cite{Ishisaki2018-wt}).
Photon pulses were analyzed using optimal (matched) filtering \citep{Szymkowiak1993-md}, and event grades were assigned based on temporal proximity \citep{Tsujimoto2017-rc}.
Only temporally isolated high-primary (Hp) events were used in the spectral analysis to ensure the best achievable energy resolution and to avoid systematic uncertainties associated with secondary or multi-pixel grades at high count rates.
According to the post-launch gain report\footnote{https://heasarc.gsfc.nasa.gov/FTP/xrism/postlaunch/gainreports/2/201060010\_resolve\_energy\_scale\_report.pdf}, the Mn~K$\alpha$ calibration line exhibits a FWHM of $4.43\pm0.03$~eV and an energy shift of $-0.0975\pm0.0113$~eV for the calibration pixel.
For science pixels, the fitted FWHM is $4.41\pm0.03$~eV with a shift of $0.0092\pm0.0143$~eV.
These values are consistent with nominal in-orbit performance, and the absolute energy-scale uncertainty is 0.3~eV \citep{Eckart2025-al, Porter2025-mo}.
Independent verification using on-board calibration data confirmed the reported gain stability.

Given the extreme brightness of GX~5--1, the instrumental and cosmic background contributes
$\ll$1\% of the total counts across the full band and $<0.1\%$ in the Fe~K region.
We therefore did not subtract background spectra.
This choice does not affect the detection or modeling of narrow features, whose statistical uncertainties dominate over any residual background contribution.
Response matrices were generated using \texttt{rslmkrmf} \citep{Eckart2018-bc} with the ``Large'' option and parameter file \texttt{xa\_rsl\_rmfparam\_20190101v006.fits}.
The line-spread function includes the Gaussian core \citep{Leutenegger2025-kt}, exponential low-energy tail, escape peaks, and Si~K$\alpha$ fluorescence.
Exposure maps were created using \texttt{xaexpmap}, and ARFs were generated with \texttt{xaarfgen}, assuming a point source at the aim point and including the closed gate-valve configuration.
Mirror calibration files (\cite{Tamura2022-sf}, \cite{Boissay-Malaquin2022-rz},  \cite{Hayashi2022-tw}, and \cite{Boissay-Malaquin2024-bz}) were applied.

%------------------------------------------------------------
\subsection{Xtend}
\label{subsec:xtend}

Xtend operated in \texttt{WINDOW2BURST} mode (0.062~s exposure every 0.5~s).
Pile-up was evaluated following \citet{Yamada2012-qx}, with modifications appropriate for Xtend, yielding an estimated pile-up fraction of $\sim3\%$ within a 20$''$ radius.
While pile-up may affect the continuum slope, narrow spectral features are not significantly distorted at this level.
To maximize photon statistics, spectra were extracted from a 60-pixel radius region (1 pixel = 1.78$''$).
Redistribution matrix files (RMFs) were generated with \texttt{xtdrmf}, including Gaussian core, exponential tail, escape peaks, and Si~K$\alpha$ fluorescence \citep{Inoue2016-kk}. 
The non-X-ray background remained below 1\% of the source flux across the full band and was not subtracted.

%------------------------------------------------------------
\subsection{Light Curves and Variability}
\label{subsec:lc}

Figure~\ref{fig:lc} presents the XRISM light curves of GX~5--1 obtained with Xtend and Resolve.
The Xtend data are shown in the 0.5--2 and 4--10~keV bands with 256~s time bins, while the Resolve data are shown in the 2--5 and 5--10~keV bands with 128~s time bins.
The source exhibits significant variability on timescales of several tens of ks, including gradual intensity changes and episodes of enhanced flux.
The hardness--intensity diagrams reveal clear spectral evolution across the observation.
In particular, Phase~B corresponds to the highest count-rate interval with relatively softer spectra, whereas Phase~C shows lower intensity accompanied by harder emission.
Based on the joint evolution of intensity and hardness, the observation was divided into three intervals (Phase~A, B, and C) for time-resolved spectroscopy.
The overall behavior resembles patterns often associated with horizontal-branch--like variability in Z sources. 
The classification of the source as being on the horizontal branch is based on its characteristic count-rate variability, spectral hardness, and comparison with previously established Z-track patterns (e.g., \cite{Kuulkers1994-zm, Asai1994-zm}).
The variability on timescales of tens of ks is comparable to secular motion along the Z track, and likely reflects gradual changes in the mass accretion rate and inner flow configuration, rather than rapid dynamical processes.

\section{Spectral Analysis}
\label{sec:analysis}

\subsection{Overview of Time-averaged and Phase-selected Spectra}

\begin{figure*}[ht]
  \begin{center}
    \includegraphics[width=\linewidth]{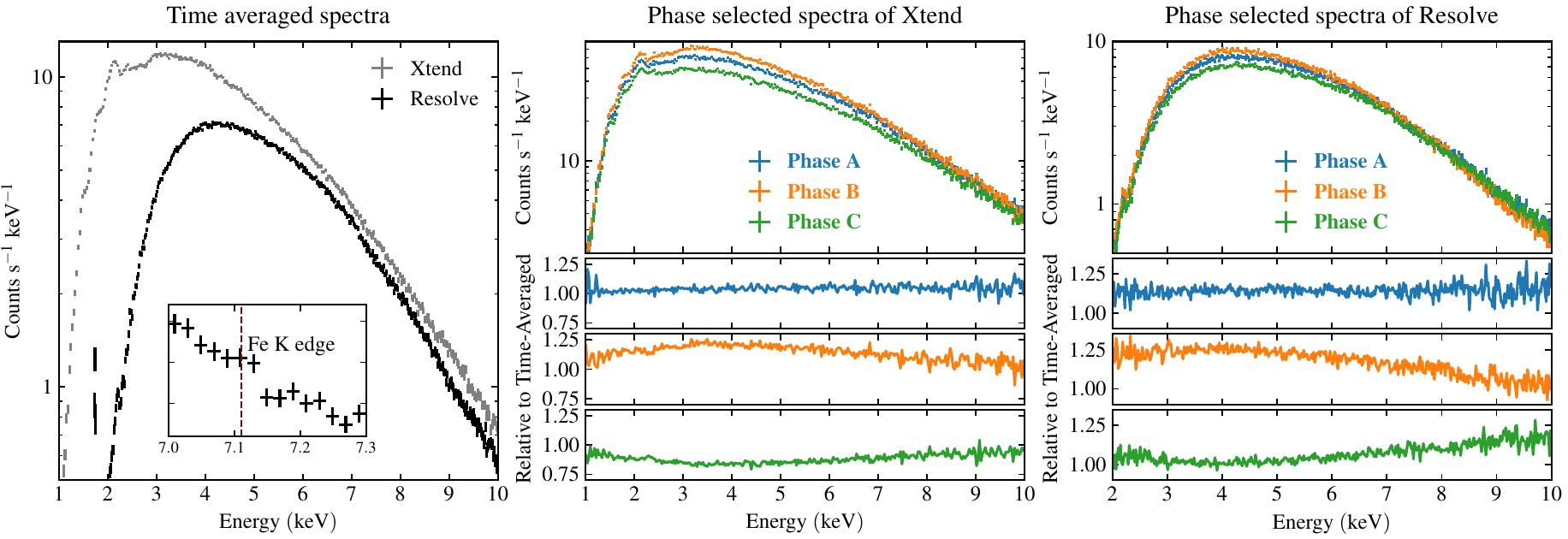} 
  \end{center}
  \caption{
  Comparison of the time-averaged and phase-selected spectra obtained withXRISM/Resolve and Xtend.
  (\textit{Left}) Time-averaged spectra of GX~5--1 in detector count space.
  The Resolve and Xtend spectra are shown in black and gray, respectively, with the Xtend spectrum scaled for visual comparison.
  The insets highlight the regions around the S~K and Fe~K absorption edges.
  (\textit{Middle}) Phase-resolved spectra obtained with Xtend for three representative intervals (Phase~A, B, and C), shown together with their ratios to the time-averaged spectrum in the lower panels.
  (\textit{Right}) Same as the middle panel but for the Resolve data.
  The ratio panels emphasize the spectral variations relative to the time-averaged spectrum, revealing systematic changes in spectral shape among the three phases.
  {Alt text: Comparison of time-averaged and phase-resolved spectra of GX 5-1 obtained with XRISM/Resolve and Xtend, including ratios to the time-averaged spectrum. The spectra show systematic changes in spectral shape across the three phases.}
  }
  \label{fig_spec}
\end{figure*}

Figure~\ref{fig_spec} presents a comparison between the time-averaged and phase-selected spectra of GX~5--1 obtained with Resolve and Xtend.
The left panel shows the time-averaged spectra in detector count space, where the Xtend spectrum is scaled for clarity.
Both instruments exhibit the characteristic broad continuum of this bright Z-source, while the insets highlight the energy regions around the Fe~K absorption edges.

The middle and right panels display spectra extracted from three representative intervals (Phase~A, B, and C) for Xtend and Resolve, respectively. In each case, the lower panels show the ratio of the phase-selected spectra to the time-averaged spectrum, which provides a direct visualization of spectral variability independent of the overall flux level. 
Both instruments show systematic spectral differences among the three phases.
In particular, the ratio plots reveal gradual changes in spectral curvature across the 2--10~keV band, indicating that the spectral shape evolves with source phase rather than simply scaling with flux.
These phase-dependent variations motivate a more detailed spectral analysis described below.

\subsection{Time-averaged Continuum Modeling}

\begin{figure*}[ht]
  \begin{center}
    \includegraphics[width=0.99\linewidth]{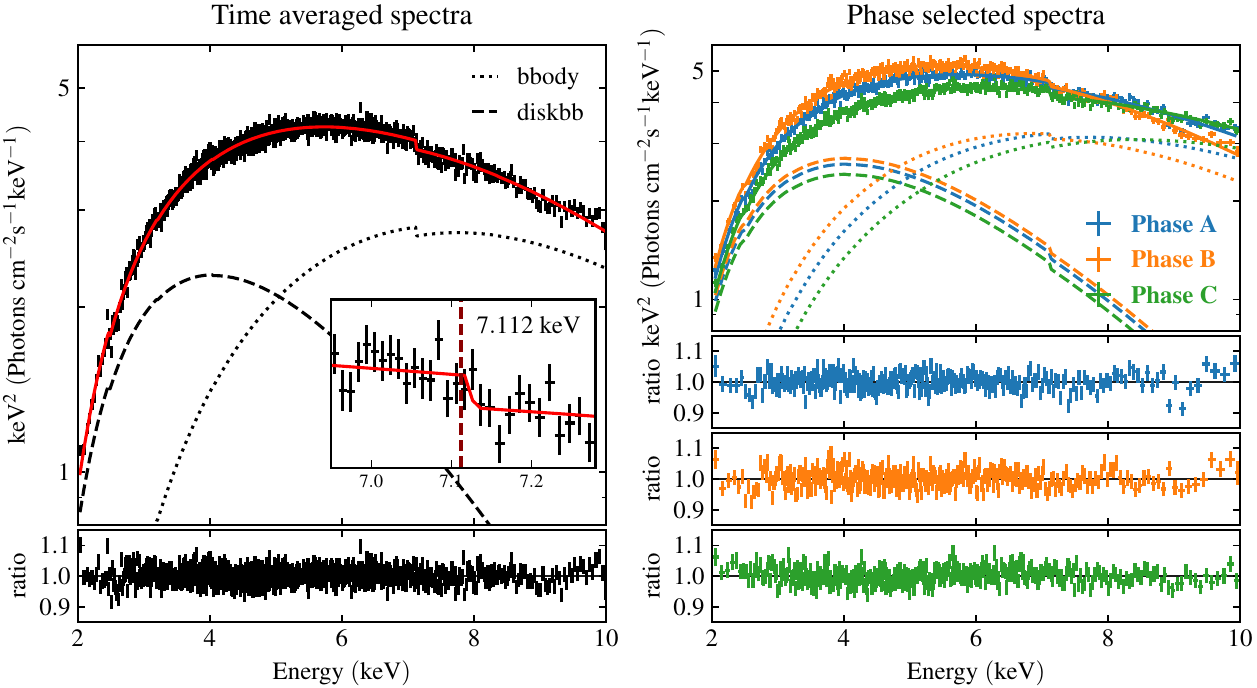} 
  \end{center}
  \caption{
  (Left) Time-averaged spectrum of GX 5--1 fitted with a two-component continuum model consisting of a multi-color disk blackbody (\texttt{diskbb}) and a single-temperature blackbody (\texttt{bbody}). The unfolded spectrum is shown in units of keV$^2$ (photons cm$^{-2}$ s$^{-1}$ keV$^{-1}$), together with the best-fit model (red solid curve). The individual model components are indicated by dashed and dash-dotted curves for \texttt{diskbb} and \texttt{bbody}, respectively. The lower panel shows the data-to-model ratio. The inset highlights the Fe K band, where the Fe-K absorption edge is visible.
  (Right) Phase-selected spectra (Phase A: blue; Phase B: orange; Phase C: green) fitted with the same \texttt{diskbb}$+$\texttt{bbody} model. The upper panel shows the unfolded spectra with the corresponding best-fit models and component contributions, while the lower three panels display the data-to-model ratios for each phase. The continuum parameters vary systematically with intensity, reflecting spectral evolution across the three phases. 
  {Alt text: Time-averaged and phase-resolved spectra of GX 5-1 fitted with a disk blackbody plus blackbody continuum model, including data-to-model ratios. The spectra show systematic changes in continuum parameters with intensity across the three phases.}
  }
\label{fig_spec_fit}
\end{figure*}

To establish a baseline spectral model, we fitted the time-averaged Resolve spectrum in the 2--10~keV band with an absorbed two-component continuum, \texttt{tbabs}$*$\texttt{(diskbb+bbody)}.
This phenomenological model represents emission from an optically thick accretion disk together with a hotter thermal component, commonly interpreted as originating from the neutron-star boundary layer.
The model provides an acceptable description of the spectrum (Table~\ref{tab:fitresults}).
The best-fit hydrogen column density is $N_{\rm H} = 2.41^{+0.17}_{-0.13} \times 10^{22}\ {\rm cm^{-2}}$, consistent with previous measurements within the uncertainties.
The best-fit temperatures are $kT_{\rm in} = 1.39^{+0.33}_{-0.18}$~keV for the disk component and $kT_{\rm bb} = 1.90^{+0.24}_{-0.10}$~keV for the blackbody component.
The fit yields $C/{\rm d.o.f.} = 3082.6/3037$, with no structured residuals exceeding a few percent across the fitted energy range.

We also fitted the spectrum with an alternative continuum model including thermal Comptonization, \texttt{tbabs*(diskbb+nthComp)}, which is commonly adopted for luminous neutron-star LMXBs (the eastern model).
The alternative model provides a fit of comparable quality to that of the baseline model, and the resulting 3$\sigma$ upper limits on the equivalent widths of narrow Fe~K features remain consistent within the statistical uncertainties.

The unfolded spectrum and residuals are shown in figure~\ref{fig_spec_fit} (left).
The residuals are smooth over the entire band and show no evidence for reflection features or discrete emission or absorption lines.
Possible weak Fe~K emission and absorption features are examined in more detail in the following subsection.
A closer inspection of the 6.8--7.5~keV band likewise reveals no persistent line-like structures.

The non-detection of Fe~K emission or absorption features is notable given the high luminosity of GX~5--1.
In many luminous neutron-star systems, reflection features are observed when the accretion disk subtends a substantial solid angle as seen from the Comptonizing region.
The present result is consistent with either a small reflection fraction, a highly ionized reflector that suppresses line contrast, or rapid spectral variability that dilutes weak features in the time-averaged spectrum.
The adopted continuum model therefore provides a suitable baseline for searching for weak spectral features with Resolve.

In particular, no sharp neutral Fe~K absorption edge is detected at 7.112~keV within the present sensitivity.
This leaves open the possibility of constraining the chemical state of interstellar iron along the line of sight, including the relative contributions of gas-phase and solid-phase components.
A detailed analysis of the Fe~K edge structure, including possible X-ray absorption fine structure (XAFS), will be presented in a separate paper.

\subsection{Time-resolved Spectroscopy and Variability}

Phase-selected spectra corresponding to the three intervals defined in figure~\ref{fig:lc} are shown in figure~\ref{fig_spec} (right).
As discussed in Appendix~\ref{sec:appendix_ztrack}, all three phases correspond to different segments within the horizontal branch (HB) identified from the long-term MAXI hardness--intensity diagram.
Clear spectral evolution accompanies the intensity changes.
Phase~B corresponds to the highest flux level and exhibits a relatively softer continuum, whereas Phase~C shows lower flux with a comparatively harder spectral shape.

To quantify this evolution, we fitted the Resolve spectra of the three phases using the same continuum model adopted for the time-averaged spectrum.
Parameters that are not expected to vary on short timescales were fixed to the time-averaged values in order to isolate the dominant variable components. 
All three phases are well described by the model (Table~\ref{tab:fitresults}).

The unfolded spectra and residuals are presented in figure~\ref{fig_spec_fit} (right).
The spectral changes are smooth and primarily reflected in systematic variations of the thermal components.
No phase requires the addition of emission or absorption lines in the Fe~K band.
Residual structures at 6--8~keV vary between phases but remain statistically consistent with fluctuations expected from counting statistics.

The absence of persistent line features in both the time-averaged and phase-resolved spectra suggests that GX~5--1 does not exhibit stable reflection signatures during this observation.
If transient line-like features occur, they must be weak and short-lived, becoming diluted when integrated over the full exposure.

\begin{table*}[htbp]
\caption{Spectral fitting results of the time-averaged and time-resolved XRISM/Resolve spectra of GX 5--1 using the model \texttt{tbabs} $*$ (\texttt{diskbb}$+$\texttt{bbody}).
The four columns correspond to the time-averaged spectrum and three representative temporal phases (A, B, and C). All uncertainties are quoted at the 90\% confidence level.}
\centering
\label{tab:fitresults}
\renewcommand{\arraystretch}{1.2}
\begin{tabular}{ccccc}
\hline
\hline
Parameter                       & Time-averaged                         & Phase A & Phase B & Phase C \\
\hline
 Exposure time (ks) & 51.1 & 13.8 & 15.9 & 21.4 \\
 $N_{H}~(\rm 10^{22}~cm^{-2})$  & $2.41^{+0.17}_{-0.13}$  & $2.41\ (\mathrm{fixed})$ & $2.41\ (\mathrm{fixed})$ & $2.41\ (\mathrm{fixed})$ \\
 $kT_{in}~\rm diskbb~(keV)$     & $1.39^{+0.33}_{-0.18} $ & $1.39\ (\mathrm{fixed})$ & $1.39\ (\mathrm{fixed})$ & $1.39\ (\mathrm{fixed})$ \\
 $R_{in}~({\rm km})^{a}$            & $13.5^{+4.1}_{-3.2} $ & $14.4\pm0.1 $ & $14.7\pm0.1 $ & $13.9\pm0.1 $\\
 $kT_{bb}~\rm bbody~(keV)$      & $1.90^{+0.24}_{-0.10}$  & $1.89\pm 0.02$ & $1.70\pm 0.02$ & $2.08\pm0.02$ \\
 $R_{bb}~({\rm km})^{a}$            & $ 6.04^{+11.29}_{-1.83} $ & $6.52\pm0.17$ & $ 8.05\pm0.19 $ & $ 5.31\pm0.11 $ \\
 $ F_{2\text{-}10~\rm keV}~({\rm 10^{-8}erg~cm^{-2}~s^{-1}}) $ & 1.00 & 1.15 & 1.21 & 1.05 \\
 $ L_{1\text{-}100~\rm keV}~({\rm 10^{38}erg~s^{-1}})^{a} $ & 1.33 & 1.51 & 1.55 & 1.44\\
 \hline
 $C\text{-}stat/dof$            & 3082.6/3037 & 2683.0/2827 & 3056.4/2870 & 2925.0/2919 \\
\hline
\hline
\end{tabular}

\begin{small}
\begin{itemize}
\setlength{\parskip}{0cm} % 
\setlength{\itemsep}{0cm} % 
\item[$^a$] $R_{\rm in}$, $R_{\rm bb}$, and $L_{1-100\,\mathrm{keV}}$ were calculated assuming a distance of $D = 9$ kpc~(\cite{Christian1997-cs}). For $R_{\rm in}$, we adopted $i=60^\circ$ (i.e., $\cos i = 0.5$). The flux values correspond to the unabsorbed flux.
\end{itemize}
\end{small}
\end{table*}

\subsection{Matched-filter Search for Narrow Features}
\label{sec:matched}

\subsubsection{Motivation for a matched-filter search}

The time-averaged spectrum of GX~5--1 can be reasonably described by a simple continuum model consisting of \texttt{diskbb + bbody} when the spectrum is grouped using an objective binning scheme such as optimal binning. Under such conditions, the residuals do not show statistically significant narrow spectral features.

\begin{figure*}[h]
  \begin{center}
    \includegraphics[width=0.9\linewidth]{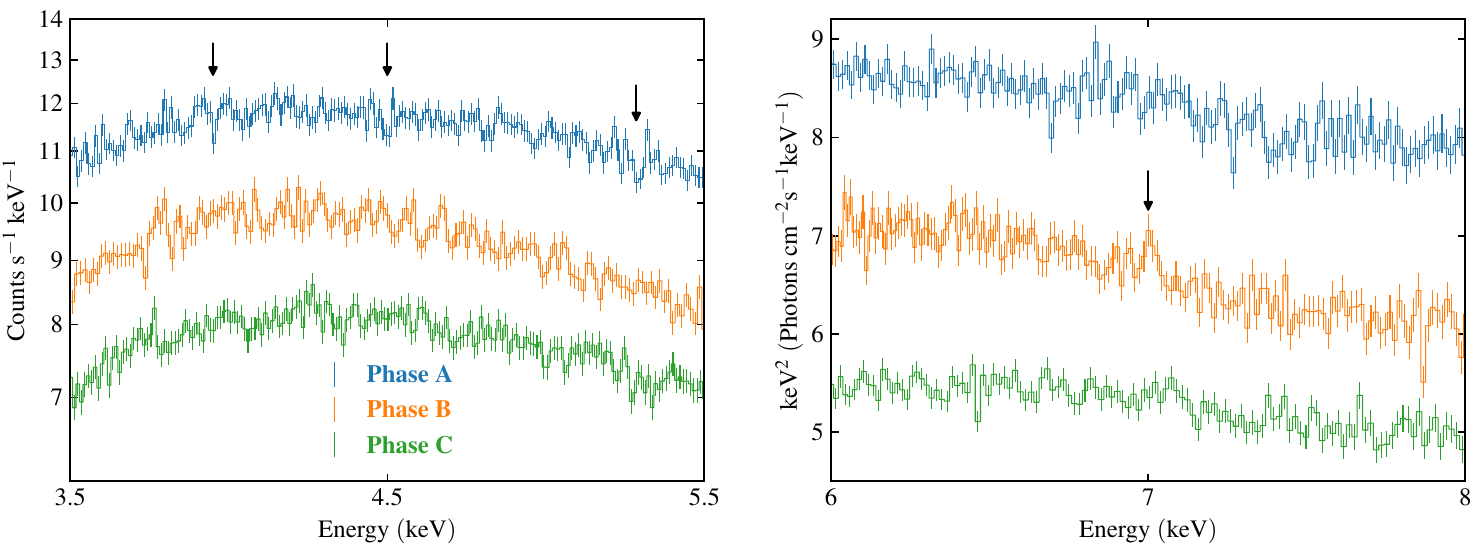} 
  \end{center}
  \caption{
  Heavily rebinned XRISM/Resolve spectra of GX~5--1 shown to illustrate possible weak structures that motivated the matched-filter search presented in this work.
  Left: spectra in detector space (counts~s$^{-1}$~keV$^{-1}$).
  Right: approximately unfolded spectra obtained using the XSPEC command
  \texttt{plot uuef}.
  The spectra corresponding to three representative intervals (Phase~A, Phase~B, and Phase~C) are shown with vertical offsets for visual clarity.
  Arrows indicate locations where small deviations from a smooth continuum may be visually suggested after heavy rebinning.
  These rebinned spectra are shown only for qualitative illustration.
  Because the appearance of such structures depends strongly on the chosen binning scheme, they are not used for quantitative spectral modeling or significance estimation.
  Instead, we perform a systematic search for weak spectral features using a matched-filter technique applied to the original high-resolution spectra (\S~\ref{subsec:matchedfilter}).
  {Alt text: Heavily rebinned XRISM/Resolve spectra of GX 5-1 for three phases, shown in detector and unfolded representations to illustrate possible weak deviations from a smooth continuum. These features are qualitative and motivated a matched-filter search on the original high-resolution spectra.}
  }
\label{fig:structure_hint}
\end{figure*}

However, when the spectrum is heavily rebinned for visualization purposes (e.g., using the XSPEC command \texttt{setplot rebin}), small deviations from a smooth continuum may appear visually.
Figure~\ref{fig:structure_hint} shows an example of such heavily rebinned spectra in both detector space and approximately unfolded space.
In these representations, weak structures around several energies may appear suggestive when the data are smoothed over large bins.

It is important to note that the visual appearance of such features depends sensitively on the adopted binning scheme and therefore cannot be used as quantitative evidence for the presence of spectral lines or edges.
In particular, the significance of any apparent structure is not well-defined when the binning is adjusted interactively.

To assess the presence of weak spectral features in a statistically well-defined manner, we therefore adopt a matched-filter technique applied to the original high-resolution spectrum.
This approach allows us to perform a systematic search for narrow spectral features over the full energy band while properly accounting for the instrumental response and statistical fluctuations.
The details of the method are described in the next subsection and the matched-filter formalism is explained in Appendix~\ref{sec:appendix_matched}. 

\subsubsection{Application of matched-filter search for Resolve data}
\label{subsec:matchedfilter}

\begin{figure*}[h]
  \begin{center}
    \includegraphics[width=0.99\linewidth]{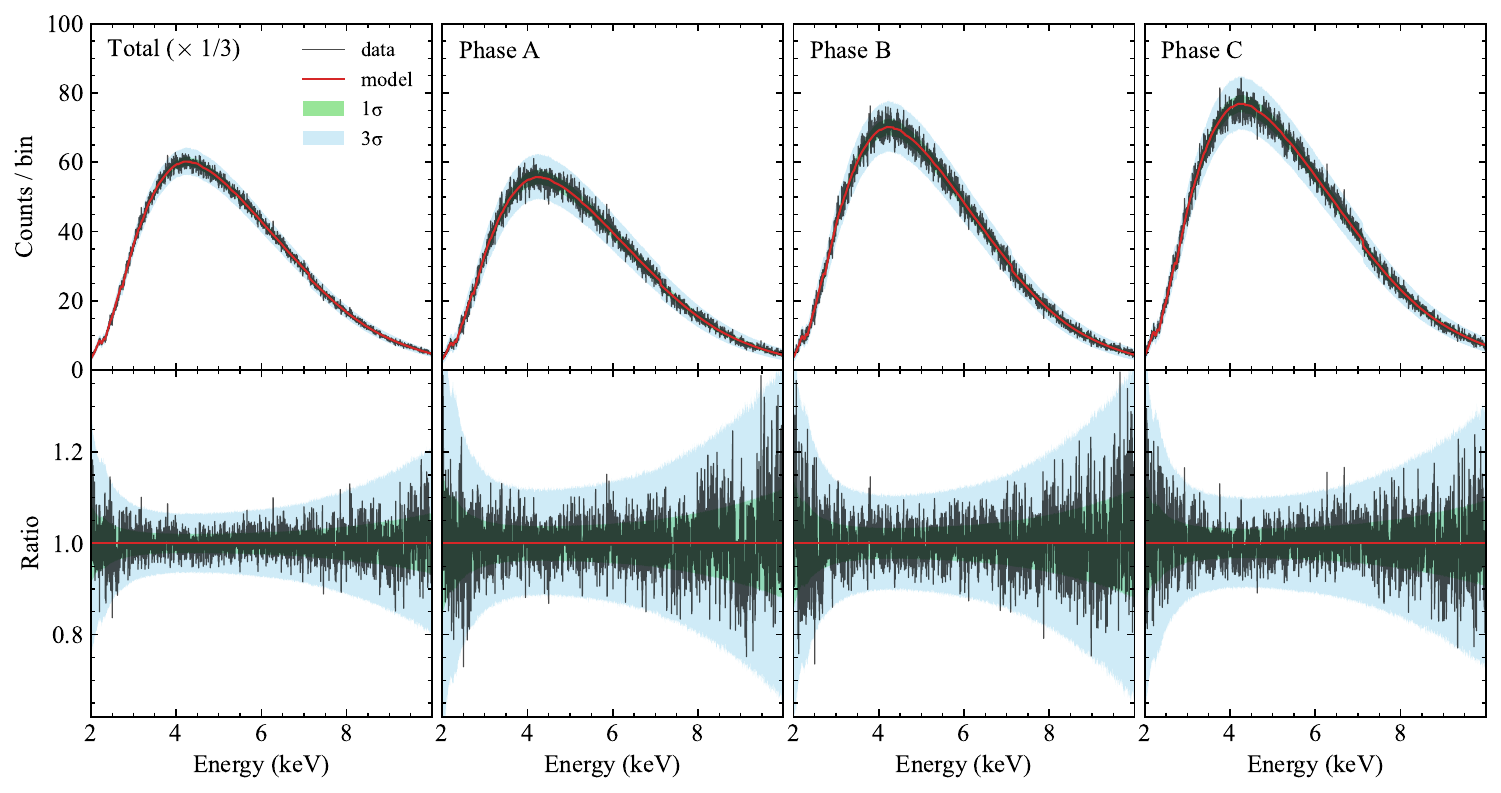}
  \end{center}
  \caption{
  Matched-filter analysis of the XRISM/Resolve spectra of GX~5--1 for the time-averaged data (leftmost panels) and three temporal segments (Phases~A, B, and C; from left to right).
  In each column, the upper panel shows the observed spectrum (black) together with the best-fitting continuum model (red), while the lower panel shows the ratio to the continuum.
  The shaded bands indicate the $\pm1\sigma$ (green) and $\pm3\sigma$ (blue) confidence intervals expected from statistical fluctuations under the continuum model, evaluated using the same matched-filter procedure.
  {Alt text: Matched-filter analysis of the time-averaged and phase-resolved XRISM/Resolve spectra of GX~5--1. The observed spectra and best-fit continuum models are shown together with data-to-model ratios. Confidence bands derived from Monte Carlo simulations indicate the expected statistical fluctuations, and no statistically significant narrow spectral features are detected.}
  }
  \label{fig_matched}
\end{figure*}

We applied a matched-filter analysis to the XRISM/Resolve spectra of GX~5--1 to search for weak and unresolved spectral features, following the methodology described in \citet{Inoue2025-dg}.
The analysis was performed for both the time-averaged spectrum and three phase-resolved spectra (Phases~A, B, and C), allowing us to examine possible temporal variations in narrow spectral structures.

In this approach, the observed spectrum is treated as a realization of a continuum model with statistical fluctuations, on top of which a weak, narrow line may be superposed.
We first determined the best-fitting continuum model for each dataset as shown in table~1,  and simulated the spectra in counts space by using \texttt{fakeit} command in \texttt{xspec}. 
A total of 10,000 simulated spectra were generated for each dataset to ensure adequate sampling of statistical fluctuations.
The matched-filter output at each trial energy $E_0$ was then evaluated by convolving the spectrum with a Gaussian kernel whose width corresponds to the instrumental energy resolution of Resolve.
This procedure effectively measures the local significance of line-like deviations from the continuum under the assumption that unresolved features are shaped by the instrumental line-spread function.

The expected statistical fluctuations of the matched-filter output were estimated using the same weighting scheme, based on the predicted continuum counts.
This allows the construction of energy-dependent confidence intervals, which we represent in terms of $\pm1\sigma$ and $\pm3\sigma$ envelopes.
To ensure adequate sampling, the trial energy was scanned with steps smaller than the instrumental resolution.

The results of this analysis are shown in figure~\ref{fig_matched}.
In both the time-averaged and phase-resolved spectra, we observe several localized deviations from the continuum at levels up to $\sim3\sigma$.
Such deviations are not unexpected in a blind search over many trial energies, and no feature is found to persist over a sufficiently broad energy range or to appear consistently across different phases.
Given that the typical width of a single spectral bin is a few eV and that the observed deviations correspond to at most $\sim10\%$ in the data-to-continuum ratio, the implied upper limits on the equivalent widths of such features are on the order of a few eV or less.

To provide quantitative constraints on representative Fe K transitions, we estimated the $3\sigma$ equivalent-width upper limits for unresolved Gaussian lines at 6.40, 6.70, and 6.97~keV, corresponding to neutral Fe K$\alpha$, Fe XXV K$\alpha$, and Fe XXVI Ly$\alpha$, respectively.
The resulting upper limits for the time-averaged and phase-resolved spectra are summarized in Table~\ref{tab:ew_limits}.

\begin{table*}[htbp]
\caption{Three-sigma upper limits on the equivalent widths (EWs$^{b}$) of representative Fe K transitions derived from the XRISM/Resolve spectra of GX 5--1. The limits are shown for the time-averaged spectrum and for the three phase-selected spectra (Phases A, B, and C).}
\label{tab:ew_limits}
\centering
\renewcommand{\arraystretch}{1.2}
\begin{tabular}{cccccc}
\hline
\hline
Line & Energy & Time-averaged & Phase A & Phase B & Phase C \\
& (keV) & (eV) & (eV) & (eV) & (eV) \\
\hline
Fe~K$\alpha$                      & 6.40 & $6.63\times10^{-1}$ & 1.39 & $8.84\times10^{-1}$ & 1.18\\
Fe~XXV~K$\alpha$                  & 6.70 & $6.34\times10^{-1}$ & 2.05 & 1.11 & 1.14\\
Fe~XXV\hspace{-1.2pt}I~Ly$\alpha$ & 6.97 & $6.30\times10^{-1}$ & 2.51 & 1.09 & 1.80\\
\hline
\hline
\end{tabular}

\begin{small}
\begin{itemize}
\setlength{\parskip}{0cm} % 
\setlength{\itemsep}{0cm} % 
\item[b] The EW upper limits were estimated using the \texttt{XSPEC} model \texttt{tbabs}(\texttt{diskbb}+\texttt{bbody}+\texttt{gaussian}), with the Gaussian line energy fixed at the corresponding transition energy and the intrinsic line width fixed at zero (unresolved line). The quoted values correspond to 3$\sigma$ (99.73\% confidence) upper limits derived from the \texttt{eqwidth} error calculation with 10,000 Monte Carlo realizations.
\end{itemize}
\end{small}
\end{table*}

Given the exposure time of the present observation, the overall behavior of the matched-filter statistic is consistent with statistical fluctuations around the continuum model.
We therefore refrain from claiming detections of significant narrow absorption or emission features.
Instead, the matched-filter analysis provides a quantitative assessment of the sensitivity to unresolved lines, which we use to constrain the presence of such features in the subsequent analysis.

%%%
%============================================================
\section{Discussion}
\label{sec:discussion}

\subsection{Constraints on Fe~K-edge Fine Structure}

GX~5--1 has long served as an important Galactic background source for studies of interstellar absorption and dust scattering, owing to its brightness and heavily absorbed line of sight.
In particular, the dust-scattered X-ray halo and the detailed structure of the silicon K edge in this source have been investigated extensively with \textit{Chandra}, establishing GX~5--1 as a valuable probe of the intervening interstellar medium rather than merely a bright Z source \citep{Smith2006-wn,Zeegers-S-T2017-ir,Clark2018-by}.
In this broader context, the present Resolve observation extends such studies into the Fe~K band, where the calorimeter resolution of XRISM provides a qualitatively new opportunity to search for subtle edge-related structure, including possible X-ray absorption fine structure (XAFS).

In the present data, however, the Fe~K edge region is well described by a smooth continuum modified by interstellar absorption, and no statistically significant substructure is detected.
The matched-filter analysis presented in subsection~\ref{sec:matched} indicates that any localized deviations from the continuum remain at the level of $\sim$1--3$\sigma$, consistent with statistical fluctuations within the current signal-to-noise ratio after accounting for the number of independent trials in the energy scan.
In this respect, our analysis follows the general statistical caution that has long been emphasized in blind line searches, namely that local excursions must be interpreted conservatively once the search over energy space is taken into account \citep{Rutledge2003-lb,Hurkett2008-gj}.
The adoption of a matched-filtering approach is also motivated by earlier work showing that this method provides a robust and sensitive framework for weak-line searches and upper-limit estimation in high-energy spectra \citep{Miyazaki2016-bp,Inoue2025-dg,Inoue2026-eg}.

We therefore interpret the present spectrum as placing quantitative constraints on the amplitude of possible fine-structure features rather than as evidence for their detection.
These limits correspond to variations at the level of a few percent of the edge depth, or equivalently to effective equivalent widths of a few eV or less.
Any Fe~K-edge fine structure along this line of sight must therefore be either intrinsically weak or effectively smeared out by the combination of multiple absorption components and the strong intrinsic continuum variability of GX~5--1.
Future high-resolution observations of brighter and more spectrally stable Galactic sources will be important for establishing whether such fine structure can be detected robustly in the Fe~K band.

%------------------------------------------------------------
\subsection{Spectral Variability along the Z Track}

GX~5--1 is one of the classical Z sources, and its correlated spectral and timing behavior along the horizontal, normal, and flaring branches has been studied for several decades.
The phenomenological framework itself traces back to the recognition of the Z-source class in low-mass X-ray binaries \citep{Hasinger1989-ab}, while a series of detailed studies of GX~5--1 established the branch-dependent nature of its continuum evolution, quasi-periodic oscillations, time lags, and rapid variability \citep{Mitsuda1989-tx,Mitsuda1991-hl,Asai1994-zm,Kuulkers1994-zm, Vaughan1994-sk,Kamado1997-zm}.
These earlier observations already showed that GX~5--1 is a source in which the dominant observable changes are closely tied to the source position along the Z track and are often governed by continuum and timing behavior rather than by persistent narrow spectral features.

More recent analyses have connected this phenomenology to changes in the inner accretion flow, the boundary layer, and the Comptonizing region.
Models for the Z-track behavior of GX~5--1 have linked the spectral and timing evolution to changes in accretion geometry and radiation pressure \citep{Jackson-N-K2009-jb,Sriram2011-mg}, while AstroSat studies have further argued for a branch-dependent evolution of the boundary layer and corona on scales of tens of kilometers, based on lag behavior and
broad-band spectral modeling \citep{P2022-hi,Shyam-Prakash2024-rg, Thomas2024-zy}.
In addition, recent work on the extended flaring branch has suggested that some of the most extreme states may involve transient recombination features and an expansion of the thermalized boundary layer rather than a simple continuation of the ordinary flaring branch \citep{Dutta2024-da}.

In this historical context, the present Resolve observation provides the first opportunity to revisit the branch-dependent behavior of GX~5--1 with calorimeter-resolution spectroscopy in the Fe~K band.
During the XRISM observation the source was located on the horizontal branch, where previous studies have already shown that strong spectral and timing changes are present.
The Resolve data reveal clear spectral variability on timescales of tens of seconds.
The dominant changes are associated with variations in the continuum spectral shape, consistent with fluctuations in the accretion flow and the neutron-star boundary layer or Comptonizing region.
This result is therefore not a departure from the established picture of GX~5--1, but rather a high-resolution spectroscopic confirmation that the continuum remains the primary carrier of the source variability.

Weak line-like deviations are occasionally seen in individual time segments, as also indicated by the matched-filter analysis, but their statistical significance remains marginal and they do not persist in the time-averaged spectrum.
Given the limited exposure and the number of independent trials in the energy scan, such deviations are consistent with statistical fluctuations.
The absence of statistically significant Fe~K emission or absorption features suggests that any reflection- or wind-related signatures in GX~5--1 are either intrinsically weak, highly ionized, or effectively diluted by rapid continuum variability.
Future XRISM observations at other positions along the Z track, particularly on the flaring and extended flaring branches, will be important for testing whether line-forming regions become more prominent under different accretion conditions.

%------------------------------------------------------------
\subsection{GX~5--1 in the Context of Luminous NS--LMXBs}

Many luminous neutron-star low-mass X-ray binaries (NS--LMXBs) observed with broad-band instruments such as \textit{NuSTAR} show prominent reflection signatures, including broad Fe~K emission lines and Compton reflection humps, which serve as useful diagnostics of the inner
accretion geometry. 
The XRISM observations of the Z-source Cygnus X-2 presented the Fe-K emission lines from the accretion disk corona and a possible detection of an ultrafast outflow\citep{Mizumoto2025-fc}. 
Against this broader class behavior, GX~5--1 has long appeared somewhat unusual.
Historically, discussions of this source have been dominated by its rich continuum and timing phenomenology, whereas strong and persistent Fe~K diagnostics have not played a comparably central role in defining its observational character \citep{Asai1994-zm,Kuulkers1994-zm,Jackson-N-K2009-jb,P2022-hi,Thomas2024-zy}.

A major step toward quantifying this peculiarity was provided by the \textit{NuSTAR} study of \citet{Homan2018-ym}, which reported no clear evidence for Fe~K emission or a Compton reflection hump despite observations covering nearly a complete Z track.
That work placed an upper limit of $\lesssim$ a few ~eV on the Fe~K equivalent width and argued that the most plausible explanation for the absence of reflection features is a highly ionized accretion disk.
The present XRISM Resolve observation strengthens and extends that conclusion.
Even with an energy resolution of $\sim$5~eV, no statistically significant narrow emission or absorption features are detected in the time-averaged spectrum, and the matched-filter analysis indicates that any unresolved features must lie below the current detection threshold of a few eV.

The astrophysical interpretation of this result is not unique, but the range of viable scenarios is now more sharply constrained.
A highly ionized inner accretion flow remains an attractive explanation, because it would naturally suppress the contrast of bound--bound transitions in the Fe~K band, in line with the interpretation proposed by \citet{Homan2018-ym}.
At the same time, geometric dilution may also play a role.
If the innermost flow is vertically extended, or if the boundary layer and Comptonizing region subtend a large solid angle as seen by the disk, then classical reflection signatures may be weakened or partially washed out.
Rapid continuum variability provides an additional mechanism by which weak line features may be diluted in time-averaged spectra.
In this respect, GX~5--1 may represent an extreme regime of luminous accretion in which the dominant observable signature is the variability of the continuum itself, while Fe~K line formation is strongly suppressed.

This conclusion also fits naturally within the longer observational history of the source.
The importance of the present XRISM result is therefore not that it reveals an entirely unexpected new behavior, but rather that it revisits a classical source with substantially improved spectral resolution and shows that the longstanding picture of GX~5--1 as a continuum-dominated Z source continues to hold even in the Fe~K band. 
In this sense, the present non-detection is astrophysically meaningful: it demonstrates that the absence of strong Fe~K diagnostics in GX~5--1 is not simply a limitation of earlier CCD-resolution observations, but instead appears to reflect the intrinsic physical state of the source.

%============================================================
\section{Summary}
\label{sec:summary}

We have presented the first high-resolution X-ray spectroscopic study of the luminous neutron-star low-mass X-ray binary GX~5--1 using the Resolve microcalorimeter onboard XRISM, based on a 50~ks AO1 observation.

The Resolve spectrum provides a detailed view of the Fe~K band, enabling a systematic search for narrow emission and absorption features as well as fine structure around the Fe~K absorption edge.
No statistically significant narrow spectral features are detected in the time-averaged spectrum.
Using a matched-filter analysis together with time-resolved spectroscopy, we quantify the sensitivity to unresolved features and derive energy-dependent upper limits at the level of a few eV.

The time-resolved analysis reveals clear spectral variability on timescales of tens of ks, primarily driven by changes in the continuum spectral shape.
Although weak line-like deviations are occasionally observed in individual intervals, their statistical significance remains marginal and they do not persist in the time-averaged spectrum.
These behaviors are consistent with statistical fluctuations once the number of search trials is taken into account.

During the XRISM observation the source was located on the horizontal branch of the Z track.
This result is consistent with the long history of GX~5--1 studies, which established the source as a classical Z source whose variability is governed primarily by branch-dependent changes in its continuum and timing properties \citep{Hasinger1989-ab,Mitsuda1991-hl,Asai1994-zm,Kuulkers1994-zm,Vaughan1994-sk,Kamado1997-zm}.
The present XRISM observation extends these studies into the regime of calorimeter-resolution spectroscopy, showing that continuum variability remains the dominant observable behavior even in the Fe~K band.

Combined with previous broad-band studies reporting weak or absent reflection signatures \citep{Homan2018-ym}, the XRISM results indicate that classical Fe~K reflection or wind features in GX~5--1 are either intrinsically weak, highly ionized, geometrically diluted, or suppressed by rapid variability.
GX~5--1 may therefore represent an extreme regime of luminous neutron-star accretion in which the Fe~K band exhibits only weak discrete spectral features even at high spectral resolution.

Recent X-ray polarization measurements with the Imaging X-ray Polarimetry Explorer (IXPE) \citep{Fabiani2024-jc} reported a moderate polarization degree with a clear energy dependence in GX~5--1.
The absence of significant Fe~K emission features and a Compton reflection hump in the present XRISM observation is consistent with the interpretation of Fabiani et al. (2024), in which the observed polarization is attributed primarily to anisotropic emission and scattering in the accretion flow rather than to strong disk reflection.

Finally, GX~5--1 has long been recognized as an important probe of interstellar absorption and dust scattering \citep{Smith2006-wn,Zeegers-S-T2017-ir,Clark2018-by}.
The present Resolve observation extends these studies into the Fe~K band and places meaningful constraints on any edge-related fine structure at the level of a few eV.
These results demonstrate that XRISM/Resolve can place stringent constraints on weak spectral features even in extremely luminous and highly variable accretion flows, while highlighting the value of high-resolution calorimeter spectroscopy for studies of both accretion physics and the interstellar medium.
%%%

\begin{ack}
This observation was made possible through the sustained efforts of the XRISM team, including contributions from detector and spacecraft engineers, as well as the mission operations team.
\end{ack}

\section*{Funding}
Part of this work was supported by Japan Society for the Promotion of Science (JSPS) KAKENHI grant Nos.\ 24K00696 (T.H.), JP24K00672 (S.Y.), 	23K22543 (S.Y.), JP21K13958 (M.M.), 
23K25882 and 23H04895 (T.M.), 22H01269 (T.K.), Yamada Science Foundation (M.M.), and 
JSPS Core-to-Core Program (grant number:JPJSCCA20220002).

\section*{Data availability} 
The XRISM data underlying this article are available in heasarc data repository (NASA/GSFC) or DARTS (JAXA/ISAS). All the data reduction tools are available as a ftools package. 

%\newpage

\appendix

\section{Identification of the Source Position on the Z Track}
\label{sec:appendix_ztrack}

\begin{figure}[th]
  \begin{center}
    \includegraphics[width=0.99\linewidth]{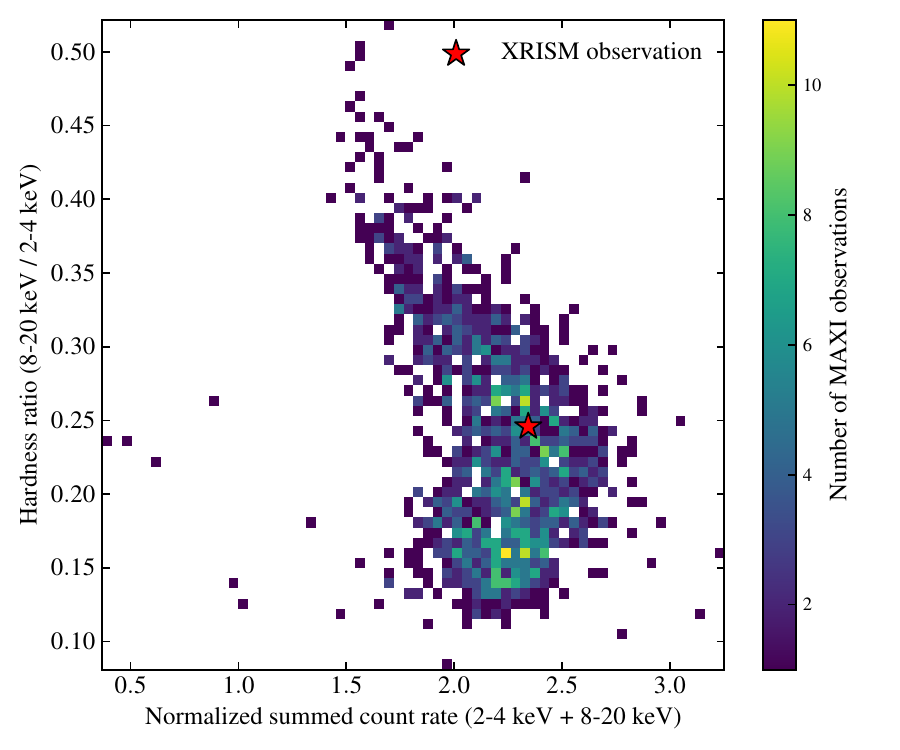}
  \end{center}
  \caption{
  Hardness--intensity diagram of GX~5--1 constructed from daily MAXI/GSC observations obtained between 2020 April 1 and 2026 April 30.
  The hardness ratio is defined as the 8--20~keV to 2--4~keV count-rate ratio, while the normalized intensity is the summed 2--4~keV and 8--20~keV count rate normalized by its long-term average.
  The color scale shows the number of MAXI observations in each bin. The red star indicates the position of the XRISM observation, which is located in the high-hardness region commonly associated with the horizontal branch.
  {Alt text: Hardness--intensity diagram of GX~5--1 constructed from long-term MAXI/GSC observations. The color scale represents the number of observations in each bin. The position of the XRISM observation is marked by a red star and lies in the high-hardness region associated with the horizontal branch.}
  }
  \label{fig:maxi_hid}
\end{figure}

To clarify the accretion state during the XRISM observation, we examined the long-term X-ray variability of GX~5--1 using publicly available data from the Gas Slit Camera (GSC) onboard the Monitor of All-sky X-ray Image (MAXI; \citep{Matsuoka2009-mx}).
We constructed a hardness--intensity diagram (HID) from daily averaged observations obtained between 2020 April 1 and 2026 April 30.
The hardness ratio was defined as the count-rate ratio between the 8--20~keV and 2--4~keV bands, while the normalized intensity was calculated from the summed 2--4~keV and 8--20~keV count rates normalized by their long-term average.

Figure~\ref{fig:maxi_hid} shows the resulting HID together with the position corresponding to the XRISM observation. The XRISM observation lies in the high-hardness region of the HID that is commonly identified with the horizontal branch (HB) in GX~5--1, supporting our identification of the source state during the observation.
Because Phases~A, B, and C were defined from different intervals within a single continuous XRISM observation, they should be regarded as different segments within the HB rather than different branches of the Z track. This is consistent with the classification adopted throughout this paper.

\section{Matched-Filter Formalism}
\label{sec:appendix_matched}

In this Appendix we summarize the matched-filter method used to identify weak narrow spectral features, with emphasis on a clear and self-contained derivation that connects its geometric interpretation with the least-squares formulation.
The goal is to explicitly show how the estimator, its statistical uncertainty, and the detection statistic arise from a single consistent framework.

We model the observed counts spectrum in bin $i$ as
\begin{equation}
D_i = C_i + A\,L_i(E_0),
\end{equation}
where $C_i$ represents the continuum model and $L_i(E_0)$ is a template describing a narrow line centered at a trial energy $E_0$.
The observed counts $D_i$ are subject to statistical fluctuations.
For Poisson-distributed counts, the variance is equal to the expectation value, and therefore we define
\begin{equation}
\sigma_i^2 = {\rm Var}(D_i) \simeq C_i,
\end{equation}
where the Gaussian approximation is valid in the high-count regime.
Under this approximation, the problem reduces to estimating the amplitude $A$ of a known template embedded in Gaussian noise.

We introduce the residual spectrum
\begin{equation}
R_i = D_i - C_i,
\end{equation}
which represents deviations from the continuum model.
The task is then to determine how much of the template $L_i(E_0)$ is contained in this residual.
Because each bin has a different statistical uncertainty, it is natural to define a noise-weighted inner product between two vectors
$X_i$ and $Y_i$ as
\begin{equation}
\langle X, Y \rangle
=
\sum_i \frac{X_i Y_i}{\sigma_i^2}.
\end{equation}
This definition assigns larger weight to bins with smaller noise and therefore reflects the statistical reliability of the data.
It also defines a vector space equipped with a noise-weighted metric.
In this space, the projection of a vector $X$ onto a vector $Y$ is given by
\begin{equation}
{\rm proj}_Y(X)
=
\frac{\langle X, Y \rangle}{\langle Y, Y \rangle}.
\end{equation}
Applying this general result to the present problem, we identify
$X_i = R_i$ and $Y_i = L_i(E_0)$, which yields
\begin{equation}
\hat{A}(E_0)
=
\frac{\langle R, L(E_0) \rangle}{\langle L(E_0), L(E_0) \rangle}
=
\frac{\sum_i \frac{L_i(E_0)}{\sigma_i^2} R_i}
{\sum_i \frac{L_i^2(E_0)}{\sigma_i^2}}.
\end{equation}
This shows that the matched filter extracts the component of the data aligned with the expected line profile, with proper weighting by the noise.

The same estimator can be derived from a statistical viewpoint by minimizing the weighted least-squares function
\begin{equation}
\chi^2(A)
=
\sum_i
\frac{\left[R_i - A\,L_i(E_0)\right]^2}{\sigma_i^2}.
\end{equation}
Taking the derivative with respect to $A$, we obtain
\begin{equation}
\frac{\partial \chi^2}{\partial A}
=
-2 \sum_i \frac{L_i(E_0)\left[R_i - A\,L_i(E_0)\right]}{\sigma_i^2}.
\end{equation}
Setting this to zero gives
\begin{equation}
\sum_i \frac{L_i(E_0) R_i}{\sigma_i^2}
-
A \sum_i \frac{L_i^2(E_0)}{\sigma_i^2}
= 0,
\end{equation}
which immediately leads to the same expression for $\hat{A}(E_0)$ as above.
Thus, the projection interpretation and the least-squares formulation are mathematically equivalent descriptions of the same estimation procedure.

The statistical uncertainty of the estimator can also be derived explicitly.
Introducing
\begin{equation}
N(E_0) \equiv \sum_i \frac{L_i^2(E_0)}{\sigma_i^2},
\end{equation}
the estimator can be written as
\begin{equation}
\hat{A}(E_0)
=
\frac{1}{N(E_0)}
\sum_i \frac{L_i(E_0)}{\sigma_i^2} R_i.
\end{equation}
Assuming that statistical fluctuations in different bins are independent, the variance of a linear combination gives
\begin{equation}
{\rm Var}[\hat{A}(E_0)]
=
\frac{1}{N(E_0)^2}
\sum_i
\left(
\frac{L_i(E_0)}{\sigma_i^2}
\right)^2
{\rm Var}(R_i).
\end{equation}
Since ${\rm Var}(R_i) = {\rm Var}(D_i) = \sigma_i^2$, this becomes
\begin{equation}
{\rm Var}[\hat{A}(E_0)]
=
\frac{1}{N(E_0)^2}
\sum_i \frac{L_i^2(E_0)}{\sigma_i^2}
=
\left(
\sum_i \frac{L_i^2(E_0)}{\sigma_i^2}
\right)^{-1}.
\end{equation}

We then define a normalized detection statistic
\begin{equation}
S(E_0)
=
\frac{\hat{A}(E_0)}
{\sqrt{{\rm Var}[\hat{A}(E_0)]}},
\end{equation}
which measures the amplitude of the candidate line in units of its statistical uncertainty. For a fixed trial energy, $S(E_0)$ approximately follows a standard normal distribution under the null hypothesis.

It is instructive to note that, in the limit where the noise is uniform (i.e., $\sigma_i = {\rm const.}$), the weighting becomes trivial and the estimator reduces to a simple inner product between the residual spectrum and the template.
This is formally identical to the optimal filtering used in XRISM/Resolve pulse analysis, where the signal amplitude is estimated by taking the inner product with an average pulse shape in the limit of negligible noise variation.
The present formulation can therefore be regarded as a direct generalization of that optimal filter to the case of energy-dependent noise.

In practice, $S(E_0)$ is evaluated over a range of trial energies.
To characterize the expected statistical fluctuations, Monte Carlo realizations of line-free spectra are generated from the best-fitting continuum model, including Poisson noise and instrumental response.
Each simulated spectrum is processed in the same way as the data, and the resulting distribution of $S(E_0)$ is used to construct confidence envelopes as a function of energy.

\section{X-ray Absorption Fine Structure; XAFS}
\label{sec:appendix_xafs}
X-ray Absorption Fine Structure (XAFS) refers to small-scale structures observed in the X-ray absorption spectrum near the absorption edge of a given element.
These structures arise from the interaction between a photoelectron and the local atomic environment.
When an X-ray photon with energy exceeding the binding energy of an inner-shell electron is absorbed, a photoelectron is emitted from the atom.
The emitted photoelectron is scattered by neighboring atoms, and the interference between the outgoing and scattered waves produces characteristic modulations in the absorption spectrum as a function of energy.
As a result, the detailed shape of the absorption edge contains information about the local atomic environment around the absorbing element.

The observed XAFS profile also depends on the depletion fraction, because XAFS arises from atoms incorporated into solid-state dust grains rather than isolated gas-phase atoms.
The absorption profile therefore depends on the relative fractions of the element in the gas and dust phases.

For iron, a large fraction is known to be depleted into dust grains, with typical depletion factors of $\sim0.90$--$0.99$.
In contrast, the sulfur depletion in the diffuse interstellar medium remains much less well constrained \citep{Savage1996,Jenkins2009}.
To illustrate the possible influence of sulfur depletion on the XAFS profile, we therefore consider two limiting cases in this work: all sulfur in the gas phase (0\% depletion) and all sulfur locked into dust grains (100\% depletion).

In astrophysics, XAFS can appear in the absorption spectra of bright X-ray sources observed through interstellar matter.
These features have been identified near the absorption edges of abundant elements such as oxygen, silicon, and iron \citep{Lee2009,Costantini2012,Zeegers2017}.
The detection of XAFS requires high spectral resolution because the fine structures around absorption edges are generally weak.
Future high-resolution X-ray missions equipped with microcalorimeter spectrometers, such as XRISM, are expected to provide improved constraints on the depletion and chemical
composition of interstellar dust and to enable detailed studies of XAFS around the Fe~K edge in heavily absorbed X-ray binaries \citep{Rogantini2020}.

GX~5--1 is one of the brightest persistent low-mass X-ray binaries in the Galaxy and shows substantial interstellar absorption along its line of sight.
Previous X-ray observations have reported a strong Fe~K absorption edge near 7.1~keV in the spectrum of GX~5--1 \citep{Ueda2005}.
Because of the limited spectral resolution of earlier instruments, the detailed structure of the absorption edge has not been clearly resolved.
High-resolution spectroscopy is therefore required to investigate possible fine structures near the Fe~K edge.

\begin{figure*}[th]
  \begin{center}
    \includegraphics[width=0.99\linewidth]{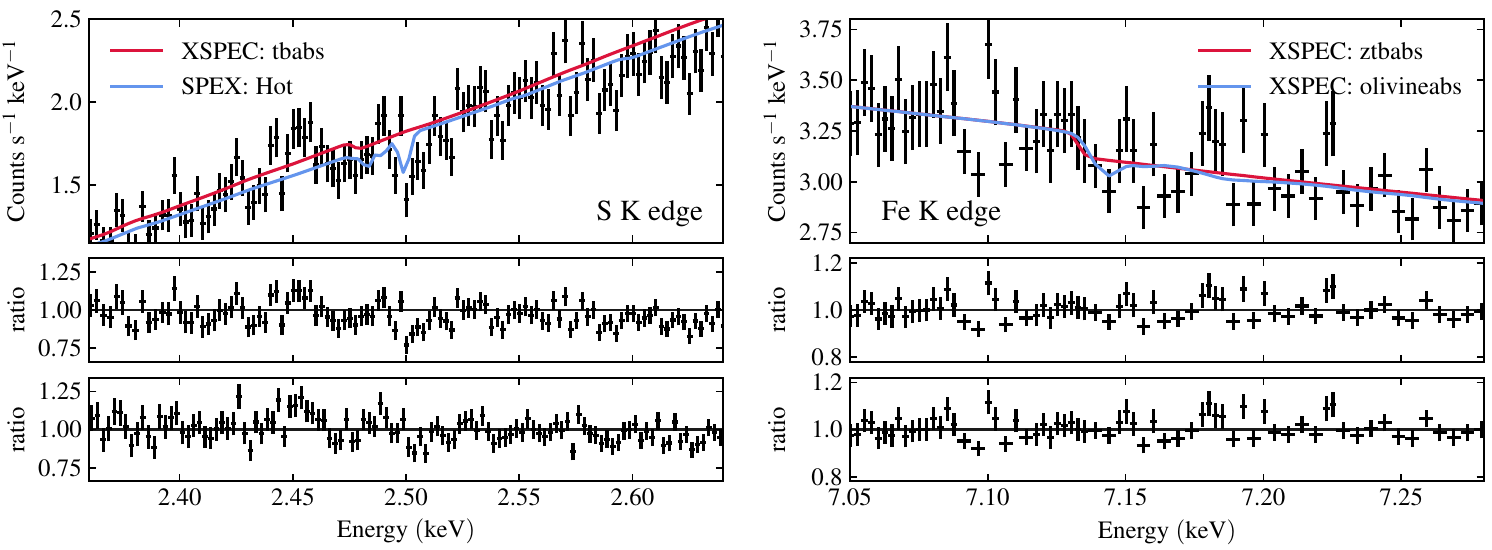}
  \end{center}
  \caption{
  Expanded views of the spectra around the S and Fe K absorption edges.
  The left panels show the spectral structures near the S K edge.
  The upper left panel presents the spectrum together with two absorption models: the \texttt{XSPEC} \texttt{tbabs} model (red solid line) and the \texttt{SPEX} \texttt{Hot} model (blue solid line).
  The middle and lower left panels show the ratios to the \texttt{tbabs} and \texttt{Hot} models, respectively.
  The right panels show the spectral structures near the Fe K edge.
  The upper right panel presents the spectrum together with two absorption models: the \texttt{XSPEC} \texttt{ztbabs} model (red solid line) and the \texttt{olivineabs} model (blue solid line).
  The middle and lower right panels show the ratios to the \texttt{ztbabs} and \texttt{olivineabs} models, respectively.
  {Alt text: XRISM/Resolve spectra around the S and Fe K absorption edges compared with different absorption models. The spectra are fitted with gas-phase and alternative absorption models, and the corresponding data-to-model ratios are shown to illustrate the agreement between the observed spectra and each model.}
  }
  \label{fig_xafs}
\end{figure*}

To illustrate the spectral features around the absorption edges, expanded views of the spectra near the S and Fe energy ranges are shown in figure~\ref{fig_xafs}.
These figures present the observed spectra in the vicinity of the expected absorption structures.
Although the overall spectral shapes are broadly consistent with previous observations, the commonly used \texttt{XSPEC} \texttt{tbabs} model does not fully reproduce the detailed structures around the S and Fe edges.

For the S~K edge, we applied the \texttt{SPEX} \texttt{Hot} model.
As shown in figure~\ref{fig_xafs}, the \texttt{Hot} model provides an alternative description of the absorption structure and better reproduces the fine-scale features than the \texttt{tbabs} model.
These results suggest that absorption models including detailed atomic and solid-state structures may provide a better description of the spectral shape around the S~K edge.

For the Fe~K edge, the observed spectral structure appears to be slightly shifted in energy relative to the prediction of the standard absorption model.
To account for this possibility, we applied the \texttt{XSPEC} \texttt{ztbabs} model, which allows the edge energy to vary through the redshift parameter.
As shown in figure~\ref{fig_xafs}, this model provides an improved description of the overall edge position compared with the fixed-energy model.
We further compared this result with the \texttt{olivineabs} model available in \texttt{XSPEC}, which incorporates more complex absorption structures associated with silicate materials.
The \texttt{olivineabs} model reproduces some of the detailed features observed near the Fe~K edge.
However, the current statistical quality of the data is insufficient to establish the presence of such detailed structures.

The adopted absorption models differ in their physical assumptions.
The \texttt{tbabs} and \texttt{ztbabs} models provide a baseline description of photoelectric absorption without solid-state XAFS structures, whereas \texttt{olivineabs} describes absorption by olivine-like silicate dust.
The present comparison should be regarded as a comparison between gas-phase and dust absorption models rather than as a measurement of the depletion fraction along the line of sight.

Another possible interpretation of structures near the Fe~K edge is mildly blueshifted Fe\,XXVI Ly$\alpha$ absorption associated with fast disk winds.
Transient disk-wind absorption in X-ray binaries can produce weak features in this energy range \citep{Miller2025-vh}.
XAFS-related features are expected to appear as relatively broad edge-like structures with fine oscillatory modulations that remain stable throughout the observation, whereas wind absorption produces narrow, velocity-shifted lines that may vary on short timescales.
High-resolution, time-resolved spectroscopy with XRISM can help distinguish between these scenarios.
No statistically significant narrow absorption features are detected in either the time-averaged or phase-resolved spectra of GX~5--1 after accounting for the number of search trials.
Weak or transient absorption features may not be evident in the time-averaged spectrum, and subtle structures near the Fe~K edge should therefore be interpreted with caution.

%%%
\bibliographystyle{aasjournal}
\bibliography{Cyg-X-1_cvt,XRISM_cvt,ASTRO-H_cvt,Calorimeter_cvt,Plasma-Physics_cvt,GX5-1_cvt,add_bib_cvt}

\begin{thebibliography}{}
\expandafter\ifx\csname natexlab\endcsname\relax\def\natexlab#1{#1}\fi
\providecommand{\url}[1]{\href{#1}{#1}}
\providecommand{\dodoi}[1]{doi:~\href{http://doi.org/#1}{\nolinkurl{#1}}}
\providecommand{\doeprint}[1]{\href{http://ascl.net/#1}{\nolinkurl{http://ascl.net/#1}}}
\providecommand{\doarXiv}[1]{\href{https://arxiv.org/abs/#1}{\nolinkurl{https://arxiv.org/abs/#1}}}

\bibitem[{Asai {et~al.}(1994)Asai, Dotani, Mitsuda, Nagase, Kamado, Kuulkers,
  \& Breedon}]{Asai1994-zm}
Asai, K., Dotani, T., Mitsuda, K., {et~al.} 1994, Publ. Astron. Soc. Jpn. Nihon
  Tenmon Gakkai, 46, 479, \dodoi{10.1093/pasj/46.5.479}

\bibitem[{Boissay-Malaquin {et~al.}(2022)Boissay-Malaquin, Hayashi, Tamura,
  Okajima, Sato, Olsen, Koenecke, Lara, Bleier, Eckart, Leutenegger, Yaqoob, \&
  Chiao}]{Boissay-Malaquin2022-rz}
Boissay-Malaquin, R., Hayashi, T., Tamura, K., {et~al.} 2022, in Space
  Telescopes and Instrumentation 2022: Ultraviolet to Gamma Ray, ed. J.-W.~A.
  den Herder, K.~Nakazawa, \& S.~Nikzad, Vol. 12181 (SPIE), 121811U,
  \dodoi{10.1117/12.2627563}

\bibitem[{Boissay-Malaquin {et~al.}(2024)Boissay-Malaquin, Hayashi, Tamura,
  Okajima, Yaqoob, Eckart, Hell, Leutenegger, Loewenstein, \&
  Sato}]{Boissay-Malaquin2024-bz}
Boissay-Malaquin, R., Hayashi, T., Tamura, K., {et~al.} 2024, in Space
  Telescopes and Instrumentation 2024: Ultraviolet to Gamma Ray, ed. J.-W.~A.
  den Herder, K.~Nakazawa, \& S.~Nikzad, Vol. 13093 (SPIE), 1309367,
  \dodoi{10.1117/12.3020154}

\bibitem[{Christian \& Swank(1997)}]{Christian1997-cs}
Christian, D.~J., \& Swank, J.~H. 1997, ApJS, 109, 177

\bibitem[{Clark(2018)}]{Clark2018-by}
Clark, G.~W. 2018, Astrophys. J., 852, 121, \dodoi{10.3847/1538-4357/aaa1f0}

\bibitem[{Costantini {et~al.}(2012)Costantini, Pinto, Kaastra,
  {et~al.}}]{Costantini2012}
Costantini, E., Pinto, C., Kaastra, J.~S., {et~al.} 2012, Astronomy \&
  Astrophysics, 539, A32, \dodoi{10.1051/0004-6361/201117520}

\bibitem[{de~Vries {et~al.}(2017)de~Vries, Haas, Yamasaki, den Herder, Paltani,
  Kilbourne, Tsujimoto, Eckart, Leutenegger, Costantini, Dercksen, Dubbeldam,
  Frericks, Laubert, van Loon, Lowes, McCalden, Porter, Ruijter, \&
  Wolfs}]{de-Vries2017-oo}
de~Vries, C.~P., Haas, D., Yamasaki, N.~Y., {et~al.} 2017, JATIS, 4, 011204,
  \dodoi{10.1117/1.JATIS.4.1.011204}

\bibitem[{Dutta {et~al.}(2024)Dutta, Pahari, Sarkar, Bhattacharyya, \&
  Bhargava}]{Dutta2024-da}
Dutta, T., Pahari, M., Sarkar, A., Bhattacharyya, S., \& Bhargava, Y. 2024,
  Monthly Notices of the Royal Astronomical Society, 535, 3383,
  \dodoi{10.1093/mnras/stae2529}

\bibitem[{Eckart {et~al.}(2018)Eckart, Adams, Boyce, Brown, Chiao, Fujimoto,
  Haas, den Herder, Hoshino, Ishisaki, Kilbourne, Kitamoto, Leutenegger,
  McCammon, Mitsuda, Porter, Sato, Sawada, Seta, Sneiderman, Szymkowiak, Takei,
  Tashiro, Tsujimoto, de~Vries, Watanabe, Yamada, \& Yamasaki}]{Eckart2018-bc}
Eckart, M.~E., Adams, J.~S., Boyce, K.~R., {et~al.} 2018, J. Astron. Telesc.
  Instrum. Syst., 4, 1, \dodoi{10.1117/1.jatis.4.2.021406}

\bibitem[{Eckart {et~al.}(2025)Eckart, Brown, Chiao, Cumbee, Fujimoto, Hell,
  Hoshino, Ishisaki, Kelley, Kenyon, Kilbourne, Kitamoto, Leutenegger, Lockard,
  Loewenstein, Magee, Miller, Mizumoto, Porter, Sato, Sawada, Shah, Shipman,
  Sneiderman, Takei, Tsujimoto, de~Vries, Watanabe, Witthoeft, Wolfs, Yamada,
  \& Yaqoob}]{Eckart2025-al}
Eckart, M.~E., Brown, G.~V., Chiao, M.~P., {et~al.} 2025, Journal of
  Astronomical Telescopes, Instruments, and Systems, 11, 042018,
  \dodoi{10.1117/1.JATIS.11.4.042018}

\bibitem[{Fabiani {et~al.}(2024)Fabiani, Capitanio, Iaria, Poutanen, Gnarini,
  Ursini, Farinelli, Bobrikova, Steiner, Svoboda, Anitra, Baglio, Carotenuto,
  Del~Santo, Ferrigno, Lewis, Russell, Russell, van~den Eijnden, Cocchi,
  Di~Marco, La~Monaca, Liu, Rankin, Weisskopf, Xie, Bianchi, Burderi, Di~Salvo,
  Egron, Illiano, Kaaret, Matt, Miku{\v{s}}incov{\'a}, Muleri, Papitto, Agudo,
  Antonelli, Bachetti, Baldini, Baumgartner, Bellazzini, Bongiorno, Bonino,
  Brez, Bucciantini, Castellano, Cavazzuti, Chen, Ciprini, Costa, De~Rosa,
  Del~Monte, Di~Gesu, Di~Lalla, Donnarumma, Doroshenko, Dov{\v{c}}iak, Ehlert,
  Enoto, Evangelista, Ferrazzoli, Garcia, Gunji, Hayashida, Heyl, Iwakiri,
  Jorstad, Karas, Kislat, Kitaguchi, Kolodziejczak, Krawczynski, Latronico,
  Liodakis, Maldera, Manfreda, Marin, Marinucci, Marscher, Marshall, Massaro,
  Mitsuishi, Mizuno, Negro, Ng, O'Dell, Omodei, Oppedisano, Pavlov, Peirson,
  Perri, Pesce-Rollins, Petrucci, Pilia, Possenti, Puccetti, Ramsey, Ratheesh,
  Roberts, Romani, Sgr{\`o}, Slane, Soffitta, Spandre, Swartz, Tamagawa,
  Tavecchio, Taverna, Tawara, Tennant, Thomas, Tombesi, Trois, Tsygankov,
  Turolla, Vink, Wu, \& Zane}]{Fabiani2024-jc}
Fabiani, S., Capitanio, F., Iaria, R., {et~al.} 2024, Astron. Astrophys., 684,
  A137, \dodoi{10.1051/0004-6361/202347374}

\bibitem[{Hasinger \& van~der Klis(1989)}]{Hasinger1989-ab}
Hasinger, G., \& van~der Klis, M. 1989, Astron. Astrophys., 225, 79

\bibitem[{Hayashi {et~al.}(2022)Hayashi, Boissay-Malaquin, Tamura, Okajima,
  Sato, Olsen, Koenecke, Lara, Bleier, Leutenegger, Eckart, Yaqoob, \&
  Chiao}]{Hayashi2022-tw}
Hayashi, T., Boissay-Malaquin, R., Tamura, K., {et~al.} 2022, in Space
  Telescopes and Instrumentation 2022: Ultraviolet to Gamma Ray, ed. J.-W.~A.
  den Herder, K.~Nakazawa, \& S.~Nikzad, Vol. 12181 (SPIE), 121815Y,
  \dodoi{10.1117/12.2627975}

\bibitem[{Homan {et~al.}(2018)Homan, Steiner, Lin, Fridriksson, A.~Remillard,
  Miller, \& Ludlam}]{Homan2018-ym}
Homan, J., Steiner, J.~F., Lin, D., {et~al.} 2018, The Astrophysical Journal,
  853, 157, \dodoi{10.3847/1538-4357/aaa439}

\bibitem[{Hurkett {et~al.}(2008)Hurkett, Vaughan, Osborne, O'Brien, Page,
  Beardmore, Godet, Burrows, Capalbi, Evans, Gehrels, Goad, Hill, Kennea,
  Mineo, Perri, \& Starling}]{Hurkett2008-gj}
Hurkett, C.~P., Vaughan, S., Osborne, J.~P., {et~al.} 2008, The Astrophysical
  Journal, 679, 587, \dodoi{10.1086/586881}

\bibitem[{Inoue {et~al.}(2025)Inoue, Enoto, Notsu, Uchida, Iwakiri, Namekata,
  \& Gendreau}]{Inoue2025-dg}
Inoue, S., Enoto, T., Notsu, Y., {et~al.} 2025, Mon. Not. R. Astron. Soc., 541,
  1403, \dodoi{10.1093/mnras/staf1048}

\bibitem[{Inoue {et~al.}(2016)Inoue, Hayashida, Katada, Nakajima, Nagino,
  Anabuki, Tsunemi, Tsuru, Tanaka, Uchida, Nobukawa, Nobukawa, Washino, Mori,
  Isoda, Sakata, Kohmura, Tamasawa, Tanno, Yoshino, Konno, \&
  Ueda}]{Inoue2016-kk}
Inoue, S., Hayashida, K., Katada, S., {et~al.} 2016, Nucl. Instrum. Methods
  Phys. Res. A, 831, 415, \dodoi{10.1016/j.nima.2016.03.071}

\bibitem[{Inoue {et~al.}(2026)Inoue, Iwakiri, Kimura, Enoto, Notsu, Uchida,
  Hamaguchi, Toriumi, Yamazaki, Tsuchiya, Murakami, Yoshioka, Arzoumanian, \&
  Gendreau}]{Inoue2026-eg}
Inoue, S., Iwakiri, W.~B., Kimura, T., {et~al.} 2026, ApJ, 1002, 65

\bibitem[{Ishisaki {et~al.}(2018)Ishisaki, Yamada, Seta, Tashiro, Takeda,
  Terada, Kato, Tsujimoto, Koyama, Mitsuda, Sawada, Boyce, Chiao, Watanabe,
  Leutenegger, Eckart, Porter, \& Kilbourne}]{Ishisaki2018-wt}
Ishisaki, Y., Yamada, S., Seta, H., {et~al.} 2018, JATIS, 4, 011217,
  \dodoi{10.1117/1.JATIS.4.1.011217}

\bibitem[{Ishisaki {et~al.}(2022)Ishisaki, Kelley, Awaki, Balleza, Barnstable,
  Bialas, Boissay-Malaquin, Brown, Canavan, Cumbee, Carnahan, Chiao, Comber,
  Costantini, den Herder, Dercksen, de~Vries, DiPirro, Eckart, Ezoe, Ferrigno,
  Fujimoto, Gorter, Graham, Grim, Hartz, Hayakawa, Hayashi, Hell, Hoshino,
  Ichinohe, Ishida, Ishikawa, James, Kenyon, Kilbourne, Kimball, Kitamoto,
  Leutenegger, Maeda, McCammon, Miko, Mizumoto, Okajima, Okamoto, Paltani,
  Porter, Sato, Sato, Sawada, Shinozaki, Shipman, Shirron, Sneiderman, Soong,
  Szymkiewicz, Szymkowiak, Takei, Tamura, Tsujimoto, Uchida, Wasserzug,
  Witthoeft, Wolfs, Yamada, \& Yasuda}]{Ishisaki2022-ci}
Ishisaki, Y., Kelley, R.~L., Awaki, H., {et~al.} 2022, in Society of
  Photo-Optical Instrumentation Engineers (SPIE) Conference Series, Vol. 12181,
  Space Telescopes and Instrumentation 2022: Ultraviolet to Gamma Ray, ed.
  J.-W.~A. den Herder, K.~Nakazawa, \& S.~Nikzad (SPIE), 121811S,
  \dodoi{10.1117/12.2630654}

\bibitem[{Ishisaki {et~al.}(2025)Ishisaki, Kelley, Awaki, Balleza, Barnstable,
  Bialas, Boissay-Malaquin, Brown, Canavan, Cumbee, Carnahan, Chiao, Comber,
  Costantini, den Herder, Dercksen, de~Vries, DiPirro, Eckart, Ezoe, Ferrigno,
  Fujimoto, Gorter, Graham, Grim, Hartz, Hayakawa, Hayashi, Hell, Hoshino,
  Ichinohe, Ishida, Ishikawa, James, Kenyon, Kilbourne, Kimball, Kitamoto,
  Leutenegger, Maeda, McCammon, Miko, Mizumoto, Noda, Okajima, Okamoto,
  Paltani, Porter, Sato, Sato, Sawada, Shinozaki, Shipman, Shirron, Sneiderman,
  Soong, Szymkiewicz, Szymkowiak, Takei, Tamura, Tsujimoto, Uchida, Wasserzug,
  Witthoeft, Wolfs, Yamada, \& Yasuda}]{Ishisaki2025-aw}
Ishisaki, Y., Kelley, R.~L., Awaki, H., {et~al.} 2025, Journal of Astronomical
  Telescopes, Instruments, and Systems, 11, 042023,
  \dodoi{10.1117/1.JATIS.11.4.042023}

\bibitem[{{Jackson, N. K.} {et~al.}(2009){Jackson, N. K.}, {Church, M. J.}, \&
  {Bałuci{\'n}ska-Church, M.}}]{Jackson-N-K2009-jb}
{Jackson, N. K.}, {Church, M. J.}, \& {Bałuci{\'n}ska-Church, M.} 2009, A\&A,
  494, 1059, \dodoi{10.1051/0004-6361:20079234}

\bibitem[{Jenkins(2009)}]{Jenkins2009}
Jenkins, E.~B. 2009, The Astrophysical Journal, 700, 1299,
  \dodoi{10.1088/0004-637X/700/2/1299}

\bibitem[{Kamado {et~al.}(1997)Kamado, Kitamoto, \& Miyamoto}]{Kamado1997-zm}
Kamado, Y., Kitamoto, S., \& Miyamoto, S. 1997, Publ. Astron. Soc. Jpn. Nihon
  Tenmon Gakkai, 49, 589, \dodoi{10.1093/pasj/49.5.589}

\bibitem[{Kelley {et~al.}(2016)Kelley, Akamatsu, Azzarello, Bialas, Boyce,
  Brown, Canavan, Chiao, Costantini, DiPirro, Eckart, Ezoe, Fujimoto, Haas, den
  Herder, Hoshino, Ishikawa, Ishisaki, Iyomoto, Kilbourne, Kimball, Kitamoto,
  Konami, Koyama, Leutenegger, McCammon, Mitsuda, Mitsuishi, Moseley, Murakami,
  Murakami, Noda, Ogawa, Ohashi, Okamoto, Ota, Paltani, Porter, Sakai, Sato,
  Sato, Sawada, Seta, Shinozaki, Shirron, Sneiderman, Sugita, Szymkowiak,
  Takei, Tamagawa, Tashiro, Terada, Tsujimoto, de~Vries, Yamada, Yamasaki, \&
  Yatsu}]{Kelley2016-xk}
Kelley, R.~L., Akamatsu, H., Azzarello, P., {et~al.} 2016, in Space Telescopes
  and Instrumentation 2016: Ultraviolet to Gamma Ray, ed. J.-W.~A. den Herder,
  T.~Takahashi, \& M.~Bautz, Vol. 9905 (SPIE), 99050V,
  \dodoi{10.1117/12.2232509}

\bibitem[{Kelley {et~al.}(2025)Kelley, Ishisaki, Costantini, Awaki, Balleza,
  Barnstable, Bialas, Boissay-Malaquin, Brown, Canavan, Carnahan, Chiao,
  Comber, Cumbee, den Herder, Dercksen, de~Vries, DiPirro, Eckart, Ezoe,
  Ferrigno, Fujimoto, Gorter, Graham, Grim, Hartz, Hayakawa, Hayashi, Hell,
  Ichinohe, Ishi, Ishida, Ishikawa, James, Kanemaru, Kenyon, Kilbourne,
  Kimball, Kitamoto, Leutenegger, Maeda, McCammon, McLaughlin, Miko, van~der
  Meer, Mizumoto, Noda, Okajima, Okamoto, Paltani, Porter, Reichenthal, Sato,
  Sato, Sato, Sawada, Shinozaki, Shipman, Shirron, Sneiderman, Soong,
  Szymkiewicz, Szymkowiak, Takei, Takeo, Tamura, Tsujimoto, Uchida, Wasserzug,
  Witthoeft, Wolfs, Yamada, Yamasaki, \& Yasuda}]{Kelley2025-pz}
Kelley, R.~L., Ishisaki, Y., Costantini, E., {et~al.} 2025, Journal of
  Astronomical Telescopes, Instruments, and Systems, 11, 042026,
  \dodoi{10.1117/1.JATIS.11.4.042026}

\bibitem[{Kuulkers {et~al.}(1994)Kuulkers, van~der Klis, Oosterbroek, Asai,
  Dotani, van Paradijs, \& Lewin}]{Kuulkers1994-zm}
Kuulkers, E., van~der Klis, M., Oosterbroek, T., {et~al.} 1994, Astron.
  Astrophys., 289, 795

\bibitem[{Lee {et~al.}(2009)Lee, Xiang, Ravel, Kortright, \&
  Flanagan}]{Lee2009}
Lee, J.~C., Xiang, J., Ravel, B., Kortright, J., \& Flanagan, K. 2009, The
  Astrophysical Journal, 702, 970, \dodoi{10.1088/0004-637X/702/2/970}

\bibitem[{Leutenegger {et~al.}(2025)Leutenegger, Brown, Chiao, Cumbee, Eckart,
  Fujimoto, Hell, Hoshino, Ishisaki, Kelley, Kenyon, Kilbourne, Kitamoto,
  Lockard, Loewenstein, Magee, Mizumoto, Porter, Sato, Sawada, Shah, Shipman,
  Sneiderman, Takei, Tsujimoto, de~Vries, Watanabe, Witthoeft, Wolfs, Yamada,
  \& Yaqoob}]{Leutenegger2025-kt}
Leutenegger, M.~A., Brown, G.~V., Chiao, M.~P., {et~al.} 2025, Journal of
  Astronomical Telescopes, Instruments, and Systems, 11, 042024,
  \dodoi{10.1117/1.JATIS.11.4.042024}

\bibitem[{Matsuoka {et~al.}(2009)}]{Matsuoka2009-mx}
Matsuoka, M., {et~al.} 2009, PASJ, 61, 999

\bibitem[{Miller {et~al.}(2006)Miller, Raymond, Fabian, Steeghs, Homan,
  Reynolds, van~der Klis, \& Wijnands}]{Miller2006-jm}
Miller, J.~M., Raymond, J., Fabian, A., {et~al.} 2006, Nature, 441, 953

\bibitem[{Miller {et~al.}(2025)Miller, Mizumoto, Shidatsu, Ballhausen, Behar,
  D{'i}az~Trigo, Done, Dotani, Garc{'i}a, Kallman, Kobayashi, Kubota, Smith,
  Takahashi, Tashiro, Ueda, Vink, Yamada, Watanabe, Iizuka, Terada, Baluta,
  Kanemaru, Ogawa, Yoshida, \& Hayashi}]{Miller2025-vh}
Miller, J.~M., Mizumoto, M., Shidatsu, M., {et~al.} 2025, Astrophysical Journal
  Letters, 988, L28, \dodoi{10.3847/2041-8213/ade25c}

\bibitem[{Mitsuda(1989)}]{Mitsuda1989-tx}
Mitsuda, K. 1989, in ESA Special Publication, Vol.~1, Two Topics in X-Ray
  Astronomy, Volume 1: X Ray Binaries. Volume 2: AGN and the X Ray Background,
  ed. J.~Hunt \& B.~Battrick, 197

\bibitem[{Mitsuda {et~al.}(1991)Mitsuda, Dotani, Yoshida, Vaughan, \&
  Norris}]{Mitsuda1991-hl}
Mitsuda, K., Dotani, T., Yoshida, A., Vaughan, B., \& Norris, J.~P. 1991, Publ.
  Astron. Soc. Jpn. Nihon Tenmon Gakkai, 43, 113, \dodoi{10.1093/pasj/43.1.113}

\bibitem[{Mitsuda {et~al.}(1984)Mitsuda, Inoue, Koyama, Makishima, Matsuoka,
  Ogawara, Shibazaki, Suzuki, Tanaka, \& Hirano}]{Mitsuda1984-ox}
Mitsuda, K., Inoue, H., Koyama, K., {et~al.} 1984, Publ. Astron. Soc. Jpn.
  Nihon Tenmon Gakkai, 36, 741, \dodoi{10.1093/pasj/36.4.741}

\bibitem[{Miyazaki {et~al.}(2016)Miyazaki, Yamada, Enoto, Axelsson, \&
  Ohashi}]{Miyazaki2016-bp}
Miyazaki, N., Yamada, S., Enoto, T., Axelsson, M., \& Ohashi, T. 2016,
  Publications of the Astronomical Society of Japan, 68, 100,
  \dodoi{10.1093/pasj/psw091}

\bibitem[{Mizumoto {et~al.}(2025)Mizumoto, Takahashi, Behar, Boissay-Malaquin,
  Corrales, Costantini, Diaz-Trigo, Miller, \& Miller}]{Mizumoto2025-fc}
Mizumoto, M., Takahashi, H., Behar, E., {et~al.} 2025, The Astrophysical
  Journal, 996, 49, \dodoi{10.3847/1538-4357/ae267b}

\bibitem[{Mori {et~al.}(2022)Mori, Tomida, Nakajima, Okajima, Noda, Tanaka,
  Uchida, Hagino, Kobayashi, Suzuki, Yoshida, Murakami, Uchiyama, Nobukawa,
  Nobukawa, Yoneyama, Matsumoto, Tsuru, Yamauchi, Hatsukade, Ishida, Maeda,
  Hayashi, Tamura, Boissay-Malaquin, Sato, Hiraga, Kohmura, Yamaoka, Dotani,
  Ozaki, Tsunemi, Kanemaru, Sato, Takaki, Terada, Miyazaki, Kusunoki, Otsuka,
  Yokosu, Yonemaru, Asahina, Asakura, Yoshimoto, Ode, Sato, Hakamata, Aoyagi,
  Aoki, Tsunomachi, Doi, Aoki, Fujisawa, Kitajima, \& Hayashida}]{Mori2022-wo}
Mori, K., Tomida, H., Nakajima, H., {et~al.} 2022, in Society of Photo-Optical
  Instrumentation Engineers (SPIE) Conference Series, Vol. 12181, Space
  Telescopes and Instrumentation 2022: Ultraviolet to Gamma Ray, ed. J.-W.~A.
  den Herder, S.~Nikzad, \& K.~Nakazawa (Proceedings of SPIE Astronomical
  Telescopes and Instrumentation 2022), 121811T, \dodoi{10.1117/12.2626894}

\bibitem[{Noda {et~al.}(2025)Noda, Mori, Tomida, Nakajima, Tanaka, Murakami,
  Uchida, Suzuki, Kobayashi, Yoneyama, Hagino, Nobukawa, Uchiyama, Nobukawa,
  Matsumoto, Tsuru, Yamauchi, Hatsukade, Odaka, Kohmura, Yamaoka, Yoshida,
  Kanemaru, Hiraga, Dotani, Ozaki, Tsunemi, Sato, Takaki, Terada, Miyazaki,
  Kusunoki, Otsuka, Yokosu, Yonemaru, Ichikawa, Nakano, Takemoto, Matsushima,
  Urase, Kurashima, Fuchi, Hayakawa, Fukuda, Kamei, Asahina, Inoue, Amano,
  Aoki, Ito, Kamatani, Takayama, Sako, Yoshimoto, Shima, Higuchi, Ninoyu, Aoki,
  Tsunomachi, \& Hayashida}]{Noda2025-aq}
Noda, H., Mori, K., Tomida, H., {et~al.} 2025, Publ. Astron. Soc. Jpn. Nihon
  Tenmon Gakkai, 77, S10, \dodoi{10.1093/pasj/psaf011}

\bibitem[{P \& Sriram(2022)}]{P2022-hi}
P, C., \& Sriram, K. 2022, Monthly Notices of the Royal Astronomical Society,
  516, 2500, \dodoi{10.1093/mnras/stac2319}

\bibitem[{Ponti {et~al.}(2012)Ponti, Fender, Begelman, Dunn, Neilsen, \&
  Coriat}]{Ponti2012-gp}
Ponti, G., Fender, R.~P., Begelman, M.~C., {et~al.} 2012, MNRAS, 422, L11

\bibitem[{Porter {et~al.}(2010)Porter, Adams, Brown, Chervenak, Chiao,
  Fujimoto, Ishisaki, Kelley, Kilbourne, McCammon, Mitsuda, Ohashi, Szymkowiak,
  Takei, Tashiro, \& Yamasaki}]{Porter2010-ae}
Porter, F.~S., Adams, J.~S., Brown, G.~V., {et~al.} 2010, in Space Telescopes
  and Instrumentation 2010: Ultraviolet to Gamma Ray, ed. M.~Arnaud, S.~S.
  Murray, \& T.~Takahashi, Vol. 7732 (SPIE), 77323J, \dodoi{10.1117/12.857888}

\bibitem[{Porter {et~al.}(2018)Porter, Boyce, Chiao, Eckart, Fujimoto,
  Ishisaki, Kilbourne, Leutenegger, McCammon, Mitsuda, Sato, Seta, Sawada,
  Sneiderman, Szymkowiak, Takei, Tashiro, Tsujimoto, Watanabe, \&
  Yamada}]{Porter2018-ip}
Porter, F.~S., Boyce, K.~R., Chiao, M.~P., {et~al.} 2018, JATIS, 4, 011218,
  \dodoi{10.1117/1.JATIS.4.1.011218}

\bibitem[{Porter {et~al.}(2025)Porter, Kilbourne, Chiao, Cumbee, Eckart,
  Fujimoto, Ishisaki, Kanemaru, Kelley, Leutenegger, Maeda, Mizumoto, Sato,
  Sawada, Sneiderman, Takei, Tsujimoto, Uchida, Watanabe, \&
  Yamada}]{Porter2025-mo}
Porter, F.~S., Kilbourne, C.~A., Chiao, M.~P., {et~al.} 2025, Journal of
  Astronomical Telescopes, Instruments, and Systems, 11, 042016,
  \dodoi{10.1117/1.JATIS.11.4.042016}

\bibitem[{Rogantini {et~al.}(2020)Rogantini, Costantini, Zeegers, Mehdipour,
  Psaradaki, Raassen, de~Vries, \& Waters}]{Rogantini2020}
Rogantini, D., Costantini, E., Zeegers, S.~T., {et~al.} 2020, Astronomy \&
  Astrophysics, 641, A149, \dodoi{10.1051/0004-6361/201936805}

\bibitem[{Rutledge \& Sako(2003)}]{Rutledge2003-lb}
Rutledge, R.~E., \& Sako, M. 2003, Monthly Notices of the Royal Astronomical
  Society, 339, 600, \dodoi{10.1046/j.1365-8711.2003.06051.x}

\bibitem[{Sato {et~al.}(2023)Sato, Uchida, \& Ishikawa}]{Sato2023-gr}
Sato, K., Uchida, Y., \& Ishikawa, K. 2023, in High-Resolution X-ray
  Spectroscopy (Singapore: Springer Nature Singapore), 93--123,
  \dodoi{10.1007/978-981-99-4409-5\_5}

\bibitem[{Savage \& Sembach(1996)}]{Savage1996}
Savage, B.~D., \& Sembach, K.~R. 1996, Annual Review of Astronomy and
  Astrophysics, 34, 279, \dodoi{10.1146/annurev.astro.34.1.279}

\bibitem[{Shipman {et~al.}(2024)Shipman, Kitamoto, Wolfs, Costantini, Eckart,
  Ferrigno, Genolet, Gorter, Grim, den Herder, Kilbourne, Leutenegger, van~der
  Meer, Mizumoto, Porter, Paltani, Sawada, Strotmann, Tsujimoto, \&
  de~Vries}]{Shipman2024-tr}
Shipman, R.~F., Kitamoto, S., Wolfs, R., {et~al.} 2024, in Space Telescopes and
  Instrumentation 2024: Ultraviolet to Gamma Ray, ed. J.-W.~A. den Herder,
  S.~Nikzad, \& K.~Nakazawa, Vol. 13093 (SPIE), 130935W,
  \dodoi{10.1117/12.3017279}

\bibitem[{Shirron {et~al.}(2018)Shirron, Kimball, James, Muench, Canavan,
  DiPirro, Bialas, Sneiderman, Boyce, Kilbourne, Porter, Fujimoto, Takei,
  Yoshida, \& Mitsuda}]{Shirron2018-wf}
Shirron, P.~J., Kimball, M.~O., James, B.~L., {et~al.} 2018, J. Astron. Telesc.
  Instrum. Syst., 4, 1, \dodoi{10.1117/1.JATIS.4.2.021403}

\bibitem[{Shyam~Prakash \& Agrawal(2024)}]{Shyam-Prakash2024-rg}
Shyam~Prakash, V.~P., \& Agrawal, V.~K. 2024, Astrophys. J., 977, 215,
  \dodoi{10.3847/1538-4357/ad93b0}

\bibitem[{Smith {et~al.}(2006)Smith, Dame, Costantini, \&
  Predehl}]{Smith2006-wn}
Smith, R.~K., Dame, T.~M., Costantini, E., \& Predehl, P. 2006, The
  Astrophysical Journal, 648, 452, \dodoi{10.1086/505650}

\bibitem[{Sriram {et~al.}(2011)Sriram, Rao, \& Choi}]{Sriram2011-mg}
Sriram, K., Rao, A.~R., \& Choi, C.~S. 2011, Astrophys. J. Lett., 743, L31,
  \dodoi{10.1088/2041-8205/743/2/L31}

\bibitem[{Szymkowiak {et~al.}(1993)Szymkowiak, Kelley, Moseley, \&
  Stahle}]{Szymkowiak1993-md}
Szymkowiak, A.~E., Kelley, R.~L., Moseley, S.~H., \& Stahle, C.~K. 1993, J. Low
  Temp. Phys., 93, 281, \dodoi{10.1007/bf00693433}

\bibitem[{Tamura {et~al.}(2022)Tamura, Hayashi, Boissay-Malaquin, Okajima,
  Sato, Olsen, Koenecke, Lara, Bleier, Eckart, Leutenegger, Yaqoob, \&
  Chiao}]{Tamura2022-sf}
Tamura, K., Hayashi, T., Boissay-Malaquin, R., {et~al.} 2022, in Space
  Telescopes and Instrumentation 2022: Ultraviolet to Gamma Ray, ed. J.-W.~A.
  den Herder, K.~Nakazawa, \& S.~Nikzad, Vol. 12181 (SPIE), 121811V,
  \dodoi{10.1117/12.2629534}

\bibitem[{Tamura {et~al.}(2024)Tamura, Hayashi, Boissay-Malaquin, Okajima,
  Sato, Eckart, Leutenegger, Yaqoob, Mori, Ishida, Maeda, Tomida, Nakajima,
  Noda, Uchida, Suzuki, Kobayashi, Yoneyama, Hagino, Nobukawa, Tanaka,
  Murakami, Uchiyama, Nobukawa, Yoshida, Matsumoto, Tsuru, Yamauchi, Hatsukade,
  Odaka, Kohmura, Yamaoka, Kanemaru, Hiraga, Dotani, Ozaki, Tsunemi, Miyazaki,
  Kusunoki, Otsuka, Yokosu, Yonematsu, Ichikawa, Nakano, Takemoto, Matsushima,
  Asahina, Fukuda, Yoshimoto, Shima, Aoyagi, Aoki, Ito, Aoki, Fujisawa,
  Shimizu, Higuchi, Simionescu, Miller, Brenneman, \&
  Hayashida}]{Tamura2024-wb}
Tamura, K., Hayashi, T., Boissay-Malaquin, R., {et~al.} 2024, in Society of
  Photo-Optical Instrumentation Engineers (SPIE) Conference Series, Vol. 13093,
  Space Telescopes and Instrumentation 2024: Ultraviolet to Gamma Ray, ed.
  J.-W.~A. den Herder, S.~Nikzad, \& K.~Nakazawa, 130931M,
  \dodoi{10.1117/12.3020109}

\bibitem[{Tashiro {et~al.}(2025)Tashiro, Kelley, Watanabe, Maejima,
  Reichenthal, Toda, Hartz, Santovincenzo, Matsushita, Yamaguchi, Petre,
  Williams, Guainazzi, Costantini, Takei, Ishisaki, Fujimoto, Henegar-Leon,
  Sneiderman, Tomida, Mori, Nakajima, Terada, Holland, Loewenstein, Miller,
  Sawada, Kallman, Kaastra, Done, Enoto, Bamba, Corrales, Ueda, Kara,
  Zhuravleva, Fujita, Arai, Audard, Awaki, Ballhausen, Baluta, Bando, Behar,
  Bialas, Boissay-Malaquin, Brenneman, Brown, Chiao, Cumbee, de Vries,
  den Herder, Díaz Trigo, DiPirro, Dotani, Carrero, Ebisawa, Eckart, Eckert,
  Eguchi, Ezoe, Ferrigno, Foster, Fukazawa, Fukushima, Furuzawa, Gallo,
  Garcia Martinez, Gorter, Grim, Gu, Hagino, Hamaguchi, Hatsukade, Hayashi,
  Hayashi, Hell, Hodges-Kluck, Horiuchi, Hornschemeier, Hoshino, Ichinohe,
  Ikuta, Iizuka, Ishi, Ishida, Ishihama, Ishikawa, Ishimura, Jaffe, Katsuda,
  Kanemaru, Kenyon, Kilbourne, Kimball, Kitamoto, Kobayashi, Kohmura, Kubota,
  Leutenegger, Maeda, Markevitch, Matsumoto, Matsuzaki, McCammon, McLaughlin,
  McNamara, Mernier, Miko, Miller, Minesugi, Mitani, Mitsuishi, Mizumoto,
  Mizuno, Mukai, Murakami, Mushotzky, Nakazawa, Natsukari, Ness, Nigo,
  Nishiyama, Nobukawa, Nobukawa, Noda, Odaka, Ogawa, Ogawa, Ogorzalek, Okajima,
  Okamoto, Ota, Ozaki, Paltani, Plucinsky, Porter, Pottschmidt, Quero, Sasaki,
  Sato, Sato, Sato, Sato, Seta, Shida, Shidatsu, Shigeto, Shipman, Shinozaki,
  Shirron, Simionescu, Smith, Soong, Suzuki, Szymkowiak, Takahashi, Takeo,
  Tamagawa, Tamura, Tanaka, Tanimoto, Terashima, Tsuboi, Tsujimoto, Tsunemi,
  Tsuru, Uchida, Uchida, Uchida, Uchiyama, Uno, Vink, Witthoeft, Wolfs, Yamada,
  Yamada, Yamaoka, Yamasaki, Yamauchi, Yamauchi, Yanagase, Yaqoob, Yasuda,
  Yoneyama, Yoshida, \& Yukita}]{Tashiro2025-ja}
Tashiro, M., Kelley, R., Watanabe, S., {et~al.} 2025, Publications of the
  Astronomical Society of Japan, 77, S1, \dodoi{10.1093/pasj/psaf023}

\bibitem[{Tashiro {et~al.}(2020)Tashiro, Maejima, Toda, Kelley, Reichenthal,
  Hartz, Petre, Williams, Guainazzi, Costantini, Fujimoto, Hayashida,
  Henegar-Leon, Holland, Ishisaki, Kilbourne, Loewenstein, Matsushita, Mori,
  Okajima, Porter, Sneiderman, Takei, Terada, Tomida, Yamaguchi, Watanabe,
  Akamatsu, Arai, Audard, Awaki, Babyk, Bamba, Bando, Behar, Bialas,
  Boissay-Malaquin, Brenneman, Brown, Canavan, Chiao, Comber, Corrales, Cumbee,
  de~Vries, den Herder, Dercksen, Diaz-Trigo, DiPirro, Done, Dotani, Ebisawa,
  Eckart, Eckert, Eguchi, Enoto, Ezoe, Ferrigno, Fujita, Fukazawa, Furuzawa,
  Gallo, Gorter, Grim, Gu, Hagino, Hamaguchi, Hatsukade, Hawthorn, Hayashi,
  Hell, Hiraga, Hodges-Kluck, Horiuchi, Hornschemeier, Hoshino, Ichinohe, Iga,
  Iizuka, Ishida, Ishihama, Ishikawa, Ishimura, Jaffe, Kaastra, Kallman, Kara,
  Katsuda, Kenyon, Kimball, Kitaguti, Kitamoto, Kobayashi, Kobayashi, Kohmura,
  Kubota, Leutenegger, Li, Lockard, Maeda, Markevitch, Martz, Matsumoto,
  Matsuzaki, McCammon, McLaughlin, McNamara, Miko, Miller, Miller, Minesugi,
  Mitani, Mitsuishi, Mizumoto, Mizuno, Mukai, Murakami, Mushotzky, Nakajima,
  Nakamura, Nakazawa, Natsukari, Nigo, Nishioka, Nobukawa, Nobukawa, Noda,
  Odaka, Ogawa, Ohashi, Ohno, Ohta, Okamoto, Ota, Ozaki, Paltani, Plucinsky,
  Pottschmidt, Sampson, Sasaki, Sato, Sato, Sato, Sawada, Seta, Shibano, Shida,
  Shidatsu, Shigeto, Shinozaki, Shirron, Simionescu, Smith, Someya, Soong,
  Sugawara, Sugawara, Szymkowiak, Takahashi, Takeshima, Tamagawa, Tamura,
  Tanaka, Tanimoto, Terashima, Tsuboi, Tsujimoto, Tsunemi, Tsuru, Uchida,
  Uchida, Uchiyama, Ueda, Uno, Vink, Watanabe, Wittheof, Wolfs, Yamada,
  Yamaoka, Yamasaki, Yamauchi, Yamauchi, Yanagase, Yaqoob, Yasuda, Yoshida,
  Yoshioka, \& Zhuravleva}]{Tashiro2020-iz}
Tashiro, M.~S., Maejima, H., Toda, K., {et~al.} 2020, in Society of
  Photo-Optical Instrumentation Engineers (SPIE) Conference Series, Vol. 11444,
  Space Telescopes and Instrumentation 2020: Ultraviolet to Gamma Ray, ed.
  J.-W.~A. den Herder, K.~Nakazawa, \& S.~Nikzad (SPIE), 1144422,
  \dodoi{10.1117/12.2565812}

\bibitem[{Thomas {et~al.}(2024)Thomas, Giridharan, Gudennavar, \&
  Bubbly}]{Thomas2024-zy}
Thomas, N.~T., Giridharan, L., Gudennavar, S.~B., \& Bubbly, S.~G. 2024, Mon.
  Not. R. Astron. Soc., 534, 3068, \dodoi{10.1093/mnras/stae2268}

\bibitem[{Tsujimoto {et~al.}(2017)Tsujimoto, Mitsuda, Kelley, den Herder,
  Bialas, Boyce, Chiao, de~Vries, DiPirro, Eckart, Ezoe, Fujimoto, Hoshino,
  Ishikawa, Ishisaki, Kilbourne, Koyama, Leutenegger, Masters, Mitsuishi, Noda,
  Okajima, Okamoto, Porter, Sato, Sato, Savinell, Sawada, Seta, Shirron,
  Sneiderman, Takei, Tamagawa, Tashiro, Watanabe, Yamada, Yamasaki, \&
  Yatsu}]{Tsujimoto2017-rc}
Tsujimoto, M., Mitsuda, K., Kelley, R.~L., {et~al.} 2017, JATIS, 4, 011205,
  \dodoi{10.1117/1.JATIS.4.1.011205}

\bibitem[{Uchida {et~al.}(2025)Uchida, Mori, Tomida, Nakajima, Noda, Tanaka,
  Murakami, Suzuki, Kobayashi, Yoneyama, Hagino, Nobukawa, Uchiyama, Nobukawa,
  Matsumoto, Tsuru, Yamauchi, Hatsukade, Odaka, Kohmura, Yamaoka, Yoshida,
  Kanemaru, Ishi, Dotani, Ozaki, Tsunemi, Miyazaki, Kusunoki, Otsuka, Yokosu,
  Yonemaru, Ichikawa, Nakano, Takemoto, Matsushima, Urase, Kurashima, Fuchi,
  Hayakawa, Fukuda, Inoue, Aoki, Takayama, Sako, Yoshimoto, Shima, Higuchi,
  Ninoyu, Aoki, Tsunomachi, Okajima, Ishida, Maeda, Hayashi, Tamura,
  Boissay-Malaquin, Sato, Takeo, Miyamoto, Matsumoto, Eckart, Hell,
  Leutenegger, \& Hayashida}]{Uchida2025-ld}
Uchida, H., Mori, K., Tomida, H., {et~al.} 2025, Publ. Astron. Soc. Jpn. Nihon
  Tenmon Gakkai, 77, S23, \dodoi{10.1093/pasj/psaf030}

\bibitem[{Ueda {et~al.}(2005)Ueda, Mitsuda, Murakami, \& Matsushita}]{Ueda2005}
Ueda, Y., Mitsuda, K., Murakami, H., \& Matsushita, K. 2005, The Astrophysical
  Journal, 620, 274, \dodoi{10.1086/426933}

\bibitem[{Ueda {et~al.}(2004)Ueda, Murakami, Yamaoka, Dotani, \&
  Ebisawa}]{Ueda2004-yg}
Ueda, Y., Murakami, H., Yamaoka, K., Dotani, T., \& Ebisawa, K. 2004, ApJ, 609,
  325

\bibitem[{Vaughan {et~al.}(1994)Vaughan, van~der Klis, Lewin, Wijers, van
  Paradijs, Dotani, \& Mitsuda}]{Vaughan1994-sk}
Vaughan, B., van~der Klis, M., Lewin, W. H.~G., {et~al.} 1994, Astrophys. J.,
  421, 738, \dodoi{10.1086/173686}

\bibitem[{Yamada {et~al.}(2012)Yamada, Uchiyama, Dotani, Tsujimoto, Katsuda,
  Makishima, Takahashi, Noda, Torii, Sakurai, Enoto, Yuasa, Koyama, \&
  Bamba}]{Yamada2012-qx}
Yamada, S., Uchiyama, H., Dotani, T., {et~al.} 2012, Publ. Astron. Soc. Jpn.
  Nihon Tenmon Gakkai, 64, 53, \dodoi{10.1093/pasj/64.3.53}

\bibitem[{Zeegers {et~al.}(2017)Zeegers, Costantini, Rogantini,
  {et~al.}}]{Zeegers2017}
Zeegers, S.~T., Costantini, E., Rogantini, D., {et~al.} 2017, Astronomy \&
  Astrophysics, 599, A117, \dodoi{10.1051/0004-6361/201629505}

\bibitem[{{Zeegers, S. T.} {et~al.}(2017){Zeegers, S. T.}, {Costantini, E.},
  {de Vries, C. P.}, {Tielens, A. G. G. M.}, {Chihara, H.}, {de Groot, F.},
  {Mutschke, H.}, {Waters, L. B. F. M.}, \& {Zeidler, S.}}]{Zeegers-S-T2017-ir}
{Zeegers, S. T.}, {Costantini, E.}, {de Vries, C. P.}, {et~al.} 2017, A\&A,
  599, A117, \dodoi{10.1051/0004-6361/201628507}

\end{thebibliography}
\end{document}